\documentclass[prl,twocolumn,showpacs]{revtex4-1}
\usepackage[margin=0.5in]{geometry}
\usepackage{hyperref}
\hypersetup{
  colorlinks   = true, 
  urlcolor     = blue, 
  linkcolor    = blue, 
  citecolor   = red 
}
\usepackage{graphicx}
\usepackage{amsmath}
\renewcommand{\vec}{\mathbf}

\begin{document}
\title{Electron-Magnon Scattering in Magnetic Heterostructures Far Out-of-Equilibrium}

\author{Erlend G. Tveten}
\affiliation{Department of Physics, Norwegian University of Science and Technology, NO-7491 Trondheim, Norway}
\author{Arne Brataas}
\affiliation{Department of Physics, Norwegian University of Science and Technology, NO-7491 Trondheim, Norway}
\author{Yaroslav Tserkovnyak}
\affiliation{Department of Physics and Astronomy, University of California, Los Angeles, California 90095, USA}
\date{\today}

\begin{abstract}
We present a theory of out-of-equilibrium ultrafast spin dynamics in magnetic heterostructures based on the $s$-$d$ model of ferromagnetism. Both in the bulk and across interfaces, the exchange processes between the itinerant $s$ and the localized $d$ electrons are described by kinetic rate equations for electron-magnon spin-flop scattering. The principal channel for dissipation of angular momentum is provided by spin relaxation of the itinerant electrons. Our theory extends interfacial spin phenomena such as torques, pumping, and the Peltier and Seebeck effects to address laser-induced rapid spin dynamics, in which the effective electron temperature may approach or even exceed the Curie temperature.
\end{abstract}


\maketitle

Controlling spin flow in magnetic heterostructures at ultrafast time scales using femtosecond (fs) laser pulses opens intriguing possibilities for spintronics~\cite{melnikovPRL11,*turgutPRL13,*choiNC14}. In these experiments the laser-induced perturbations~\cite{RevModPhys.82.2731} stir up the most extreme regime of spin dynamics, which is governed by the highest energy scale associated with magnetic order: The microscopic spin exchange that controls the ordering temperature ($T_{C}$). In contrast, at microwave frequencies the ferromagnetic dynamics in the bulk are well described by the Landau-Lifshitz-Gilbert (LLG) phenomenology~\cite{landauBOOKv9,*gilbertIEEEM04}, which has been successfully applied to the problem of the ferromagnetic resonance (FMR)~\cite{PhysRev.73.155}. At finite temperatures below $T_{C}$, the spin Seebeck and Peltier effects~\cite{uchidaNATM10,*xiaoPRB10,bauerNATM12} describe the coupled spin and heat currents across interfaces in magnetic heterostructures. These interfacial effects are governed by thermally activated magnetic degrees of freedom with characteristic frequencies that are much higher than the FMR frequency. Despite their different appearances, the microwave, thermal, and ultrafast spin dynamics are all rooted in the exchange interactions between electrons. It is thus natural to try to advance a microscopic understanding of the ultrafast spin dynamics based on the established phenomena at lower energies. 

Although some attempts have been made~\cite{koopmansPRL05, PhysRevLett.101.237401} to extend the LLG phenomenology to describe ultrafast demagnetization in bulk ferromagnets, no firm connection exists between the ultrafast spin generation at interfaces and the microwave spin-transfer and spin-pumping effects~\cite{Brataas:2012fk} or the thermal spin Seebeck and Peltier effects. In this Letter, we unify the energy regimes of microwave, thermal, and ultrafast spin dynamics in magnetic heterostructures from a common microscopic point of view. In the ultrafast regime, rapid heating of itinerant electrons leads to demagnetization of localized spins via electron-magnon spin-flop scattering. The parameters that control the high and low energy limits of spin relaxation originate from the same electron-magnon interactions. In addition to the unified framework, this Letter's unique contributions are the history-dependent, non-thermalized magnon distribution function and the crucial role of the out-of-equilibrium spin accumulation among itinerant electrons as the bottleneck that limits the dissipation of spin angular momentum from the combined electronic system. 

The first reports on ultrafast demagnetization in Ni~\cite{beaurepairePRL96} challenged the conventional view of low-frequency magnetization dynamics at temperatures well below $T_C$. A multitude of mechanisms and scenarios have been suggested to explain the observed quenching of the magnetic moment. Some advocate direct coherent spin transfer induced by the irradiating laser light as the source of demagnetization~\cite{zielPRL65, *zhangPRL00, *bigotNATP09, *kirilyukRMP10}. Alternative theories argue that ultrafast spin dynamics arise indirectly through incoherent heat transfer to the electron system~\cite{koopmansNATM10, *schellekensPRB13,mentinkPRL12}. Recent experiments have demonstrated that non-local laser irradiation also induces ultrafast demagnetization~\cite{eschenlohrNATM13}, and atomistic modeling~\cite{Ostler:2012sf} supports the view that heating of magnetic materials is sufficient to induce ultrafast spin dynamics. Although it is unlikely that a single explanation applies to all experimental scenarios of ultrafast demagnetization, it may be possible to address many measurements of different magnetic materials from a common microscopic view. 

THz magnon excitations in metallic ferromagnets have recently been proposed as an important element of ultrafast demagnetization by several authors~\cite{illgPRB13, zhangPRL12rt}. The elementary interaction that describes these excitations is the electron-magnon scattering. Our theory is based on kinetic equations for the low-frequency spin and charge transport associated with the microwave magnetization dynamics in heterostructures~\cite{brataasPRL00,*tserkovPRL02sp,*tserkovRMP05} and with the linear spin-caloritronic response~\cite{bauerNATM12,hoffmanPRB13}. We extend these theories to treat high-temperature far-from-equilibrium spin dynamics, in which transport is dominated by magnons and hot electrons. Electron-magnon scattering plays a critical role in this regime. We base our understanding of this interaction on the \textit{transverse spin diffusion}~\cite{tserkovPRB09sd} in the bulk and the \textit{spin-mixing physics}, e.g.,~spin transfer and spin pumping~\cite{brataasPRL00,benderPRL12,*benderPRB14}, at the interfaces.  

In our approach, we assume that the localized spins that result in macroscopic magnetization are distinct from the itinerant electron bath at the energy scales of interest. In turn, the electron bath acts as a reservoir for absorbing and dissipating spin angular momentum. The experimental techniques that are used to measure ultrafast spin dynamics typically probe the absorption edge of localized electron orbitals buried deeply below the Fermi level~\cite{vorakiatPRX12}. Therefore, these methods may be less sensitive to the itinerant electron spin density deposited in the vicinity of the Fermi energy, such that the electron reservoir becomes partially invisible, or "dark," to these element-specific probes. We assume that the relaxation of the spin accumulation among the itinerant electrons is considerably slower than the spin transfer from the localized to the itinerant electrons. In this way we avoid introducing any novel ultrafast channels for dissipation of spin angular momentum.

According to the accepted description of relaxation in ferromagnetic metals, the loss of energy and angular momentum from localized $d$ electrons is mediated by the exchange interaction to the itinerant $s$ electrons. The spin transfer from $d$ to $s$ states is accompanied by the relaxation of the $s$ electron spins to the lattice through an incoherent spin-flip process caused by the spin-orbit coupling. Eventually, most of the heat stored in the itinerant electron system escapes into lattice vibrations (phonons). Mitchell formulated such a model several decades ago to describe the longitudinal relaxation of ferromagnetic metals~\cite{mitchellPR57}. A similar description was later employed to describe Gilbert damping in ferromagnets at low frequencies~\cite{heinrichPSS67,tserkovAPL04}.

We proceed by outlining the basic formalism for ferromagnetic metals in the bulk, and we will later extend the model to describe the spin flow across a ferromagnet (F) $|$ normal-metal (N) interface, which is important for ultrafast spin dynamics in magnetic heterostructures~\cite{turgutPRL13,choiNC14}. The Hamiltonian that describes F is
\begin{equation}
\label{eq:hamiltonian}
\hat{H}=\hat{H}_0 +\hat{H}_{sd}\,,
\end{equation}
where $\hat{H}_{0}$ consists of decoupled $s$- and $d$-electron energies, including the kinetic energy of the itinerant electron bath, the $d$-$d$ exchange energy, dipolar interactions, and the crystalline and Zeeman fields. The $s$-$d$ interaction is
\begin{equation}
\label{eq:hsd}
\hat{H}_{sd} =J_{sd}\sum_{j}\vec{S}_{j}^{d}\cdot \vec{s}(\vec{r}_{j})\,,
\end{equation}
where $J_{sd}$ is the exchange energy and $\vec{S}_{j}^{d}$ [$\vec{s}(\vec{r}_{j})$] is the $d$-electron ($s$-electron) spin vector (spin density) at lattice point $j$.  We express the $s$-$d$ interaction in terms of bosonic and fermionic creation and annihilation operators:
\begin{equation}
\label{eq:hsd2nd}
\hat{H}_{sd}=\sum_{qkk'}V_{qkk'} a_{q}c_{k\uparrow}^{\dagger}c_{k'\downarrow}+{\rm H.c.}\,,
\end{equation}
where $a_{q}^{\dagger}$ ($a_{q}$) is the Holstein-Primakoff creation (annihilation) operator for magnons with wavenumber $q$ and $c_{k\sigma}^{\dagger}$ ($c_{k\sigma}$) is the creation (annihilation) operator for $s$ electrons with momentum $k$ and spin $\sigma$. $\hat{H}_{sd}$ describes how an electron flips its spin while creating or annihilating a magnon with momentum $q$ and spin $\hbar$. The scattering strength is determined by the matrix element $V_{qkk'}$. 

In Eq.~\eqref{eq:hsd2nd}, we have disregarded terms of the form $\sim a_{q}^{\dagger}a_{q'}c_{k\sigma}^{\dagger}c_{k'\sigma}$, which describe multiple-magnon scattering and do not contribute to a net change in magnetization along the spin-quantization axis. We have also disregarded higher-order terms associated with the Holstein-Primakoff expansion. When the $s$-$d$ coupling \eqref{eq:hsd} is not the dominant contribution to $\hat{H}$, we follow a mean-field approach and use Fermi's Golden Rule to compute the spin transfer between the $s$ and $d$ subsystems. We assume that all relevant energy scales are much smaller than the Fermi energy $\epsilon_F\equiv k_BT_F$ of the itinerant $s$ electrons. In this limit, the electronic continuum remains largely degenerate, with electron-hole pairs present predominantly in the vicinity of the Fermi level.

We orient the coordinate system such that the localized spin density points in the negative $z$ direction at equilibrium, with saturation value $S$ (in units of $\hbar$) in the ground state. In the presence of a magnon density $n_d$, the longitudinal spin density becomes $S_z=n_{d}-S$. The magnons are assumed to follow a quadratic dispersion relation $\epsilon_{q}=\hbar\omega_{q}=\epsilon_{0}+Aq^{2}$, where $\epsilon_{0}$ is the magnon gap and $A$ parameterizes the stiffness of the ferromagnet. $\left\langle a^{\dagger}_{q}a_{q'}\right\rangle=n(\epsilon_{q})\delta_{qq'}$ defines the magnon distribution function $n(\epsilon_{q})$, which is related to the total magnon density through $n_{d}=\int_{\epsilon_{0}}^{\epsilon_{b}} d\epsilon_{q}\mathcal{D}(\epsilon_{q})n(\epsilon_{q})$, where $\mathcal{D}(\epsilon_{q})=\sqrt{\epsilon_{q}-\epsilon_{0}}/(4\pi^{2}A^{3/2})$ is the magnon density of states. The integral over $\mathcal{D}(\epsilon_{q})$ is cut off at an energy corresponding to the bandwidth, $\epsilon_{b}\sim k_{B}T_{C}$, which is the magnon energy at the edge of the Brillouin zone.

\begin{figure}
\centering
\includegraphics[scale=0.35]{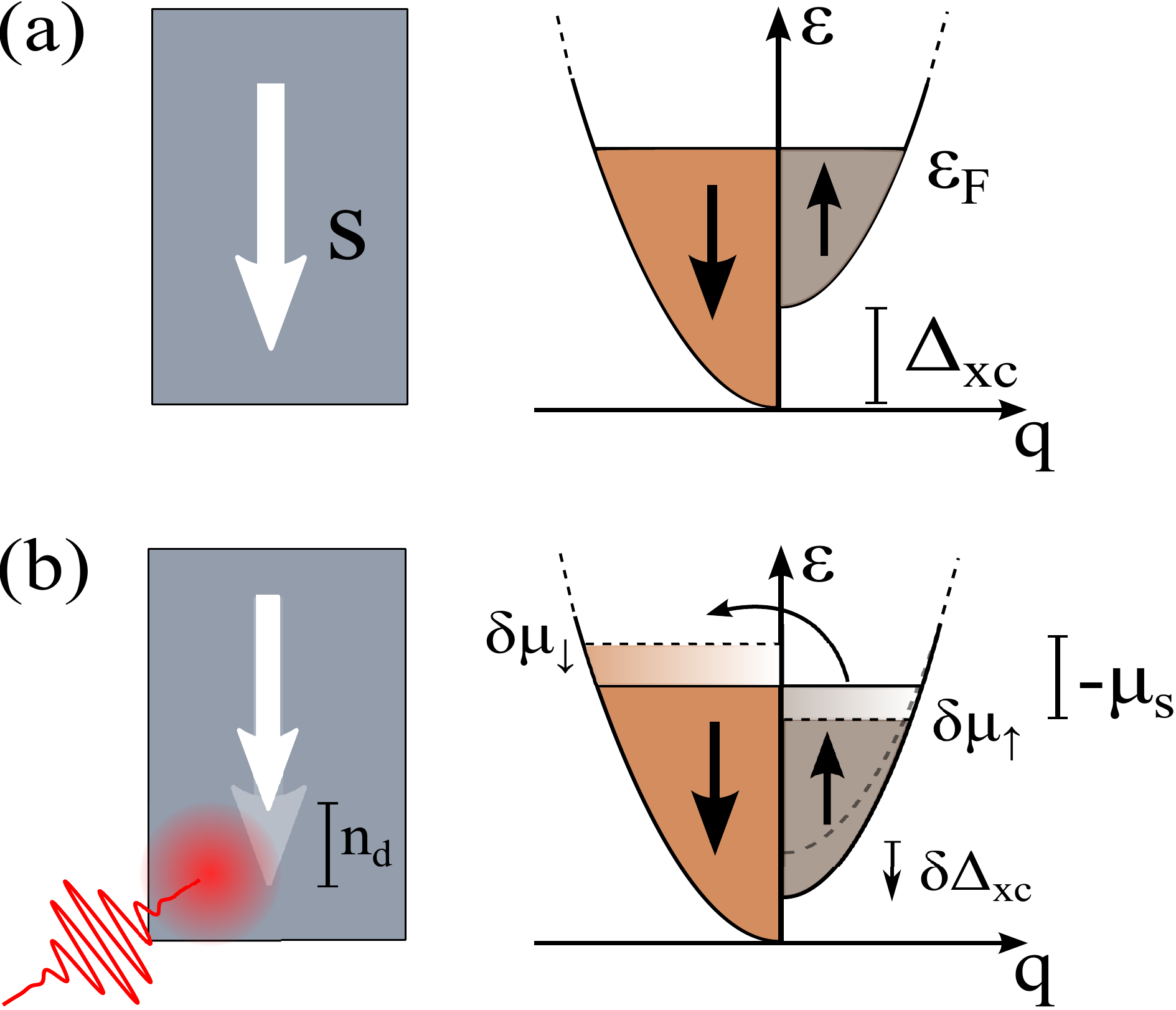}
\caption{(\textbf{a}) Sketch of the density of $s$ electron states in a ferromagnetic metal with saturation spin density $S$. At equilibrium, the exchange splitting $\Delta_{xc}$ shifts the bands for spin-up and spin-down electrons. (\textbf{b}) A pulsed laser heats the $s$ electron bath. The out-of-equilibrium spin accumulation $\mu_{s}=\delta\mu_{\uparrow}-\delta\mu_{\downarrow}$ results from two different mechanisms: 1) electron-magnon scattering induces an out-of-equilibrium spin density among the $s$ electrons, and 2) the mean-field exchange splitting is shifted by $\delta\Delta_{xc}$ by the induced non-equilibrium spin density $n_{d}$.}
\label{fig:electronSpin}
\end{figure}

Because of the $s$-$d$ exchange interaction \eqref{eq:hsd}, the itinerant $s$ electrons have a finite spin density at equilibrium; see Fig.~\ref{fig:electronSpin}. One of the key driving forces of the out-of-equilibrium spin dynamics is the spin accumulation $\mu_{s}\equiv\delta \mu_{\uparrow}-\delta\mu_{\downarrow}$. The bands for spin-up and spin-down electrons are split by $\Delta_{xc}\sim J_{sd}Sa^3$, where $a$ is the lattice constant of F. By introducing a dynamic exchange splitting, we can write $\mu_{s}=\delta n_{s}/D+\delta\Delta_{xc}$~\cite{PhysRevLett.111.167204}, where $\delta n_{s}$ is the out-of-equilibrium spin density of the $s$ electrons, $D=2D_{\uparrow}D_{\downarrow}/(D_{\uparrow}+D_{\downarrow})$, and $D_{\uparrow (\downarrow)}$ is the density of states for spin-up (spin-down) electrons at the Fermi level. Because the mean-field band splitting due to the $s$-$d$ exchange vanishes when the $d$ orbitals are fully depolarized, $\delta\Delta_{xc}/\Delta_{xc}=\pm n_{d}/S$, where the sign determines whether the $s$ and $d$ orbitals couple ferromagnetically ($-$) or antiferromagnetically ($+$).

The rate of spin-transfer (per unit volume) between the $s$ and $d$ subsystems due to electron-magnon spin-flop processes is determined from Eq.~(\ref{eq:hsd2nd}) by Fermi's Golden Rule~\cite{benderPRL12}:
\begin{equation}
\label{eq:sdcurrent}
I_{sd}=\int_{\epsilon_0}^{\epsilon_{b}}d\epsilon_{q} \Gamma (\epsilon_{q}) (\epsilon_{q}-\mu_{s})\mathcal{D}(\epsilon_{q})\left[n_{\rm BE}(\epsilon_{q}-\mu_s)-n(\epsilon_{q})\right],
\end{equation}
where $\Gamma (\epsilon_{q})$ paremeterizes the scattering rate at energy $\epsilon_{q}$. When the kinetic energy of the itinerant electrons equilibrates rapidly due to Coulombic scattering, $n_{\rm BE}(\epsilon_{q}-\mu_s)=\{\mathrm{exp}[\beta_{\mathrm{eff}}(\epsilon_{q}-\mu_s)]-1\}^{-1}$; this is the Bose-Einstein distribution function for the spin-polarized electron-hole pairs at the effective temperature $T_{\mathrm{eff}}=1/(k_B\beta_{\mathrm{eff}})$. 

However, when the time scale of the $s$-$d$ scattering is faster than the typical rates associated with magnon-magnon interactions, magnons are \textit{not} internally equilibrated shortly after rapid heating of the electron bath, as also predicted by atomistic modeling~\cite{Barker:2013kx}. Consequently, the occupation of the magnon states could deviate significantly from the thermalized Bose-Einstein distribution on the time scale of the demagnetization process. Our treatment of this central aspect differs from that of Ref.~\cite{manchonPRB12}, in which the excited magnons are assumed to be instantly thermalized with an effective spin temperature and zero chemical potential and the thermally activated electron bath is assumed to be unpolarized.

Neglecting any direct relaxation of magnons to the static lattice or its vibrations (i.e.,~phonons), $\partial_{t} n_{d}=I_{sd}/\hbar$. The equations of motion for the $s$-electron spin accumulation and the $d$-electron magnon distribution function are
\begin{eqnarray}
\label{eq:museq}
\partial_t \mu_{s} & = & -\frac{\mu_{s}}{\tau_s}+\frac{\rho}{\hbar} I_{sd}\,, \\
\label{eq:neq}
\partial_t n(\epsilon_{q}) & = &\frac{\Gamma (\epsilon_{q})}{\hbar}(\epsilon_{q}-\mu_{s})\left[n_{\rm BE}(\epsilon_{q}-\mu_s)-n(\epsilon_{q})\right]\,,
\end{eqnarray}
where $\rho$ determines the feedback of the demagnetization on $\mu_{s}$ and $\tau_{s}$ is the spin-orbit relaxation time for the $s$-electron spin density relaxing to the lattice. $\tau_{s}$ is typically on the order of picoseconds~\cite{meserveyPRL78}, and here, it represents the main channel for the dissipation of angular momentum out of the combined electronic system. In general, $\tau_{s}$ also depends on the kinetic energy of the hot electrons after laser-pulse excitation. This dependence will necessarily influence a full description of ultrafast demagnetization scenarios. However, this discussion is beyond the scope of this Letter, and we assume that $\tau_{s}$ is –independent of energy. $\rho=\rho_{D}+\rho_{\Delta}=-1/D\pm \Delta_{xc}/S$ includes effects arising from both the out-of-equilibrium spin density and the dynamic exchange splitting. For ferromagnetic ($-$) $s$-$d$ coupling, these effects add up, whereas for antiferromagnetic ($+$) coupling, they compete. When $1/D=\Delta_{xc}/S$, $\rho$ vanishes and $\mu_{s}$ decouples from the dynamics of $n_{d}$.  

The $s$-$d$ scattering rate can be phenomenologically expanded as $\Gamma(\epsilon_{q})=\Gamma_{0}+\chi(\epsilon_{q}-\epsilon_{0})$, where $\Gamma_{0}$ (which vanishes in the simplest Stoner limit \cite{tserkovPRB09sd}) parameterizes the scattering rate of the long-wavelength magnons and $\chi(\epsilon_{q}-\epsilon_{0})\propto q^{2}$ describes the enhanced scattering of higher-energy magnons due to transverse spin diffusion \cite{tserkovPRB09sd}. In general, one might expect other terms of higher order in $q$ to be present in this expansion as well. We will, however, limit ourselves to extrapolating $\Gamma(\epsilon_{q})$, which is linear in $\epsilon_{q}$, up to the bandwidth $\epsilon_{b}$, which should be sufficient for qualitative purposes.

At low temperatures, low-frequency excitations keep the spin system close to a local equilibrium, resulting in purely transverse dynamics. In the classical picture of rigid magnetic precession, the transverse relaxation time $\tau_{2}$ is determined by the longitudinal relaxation time $\tau_{1}$ as follows: $1/\tau_{2}=1/(2\tau_{1})=\alpha\omega$, where $\alpha$ is the Gilbert damping parameter and $\omega$ is the precession frequency. Indeed, in the limit $(\omega,T)\to0$, Eq.~(\ref{eq:sdcurrent}) yields
\begin{equation}
\label{eq:spinlowT}
\partial_tn_d\to-\frac{\Gamma_0}{\hbar}\epsilon_0n_d\,,
\end{equation}
which is identical to the LLG phenomenology, indicating that $\epsilon_0=\hbar\omega$ and, thus, $\Gamma_0=2\alpha$. This result establishes the important link between the scattering rate $\Gamma_{0}$ in the present treatment and the Gilbert damping parameter that is accessible through FMR experiments. We note that, according to Eqs.~\eqref{eq:sdcurrent}-\eqref{eq:neq}, the itinerant electrons exert no feedback on the $d$ electron dynamics in the adiabatic limit $(\omega,T)\to0$ as long as $\tau_s>0$. 

\begin{figure}
\centering
\includegraphics[scale=0.355]{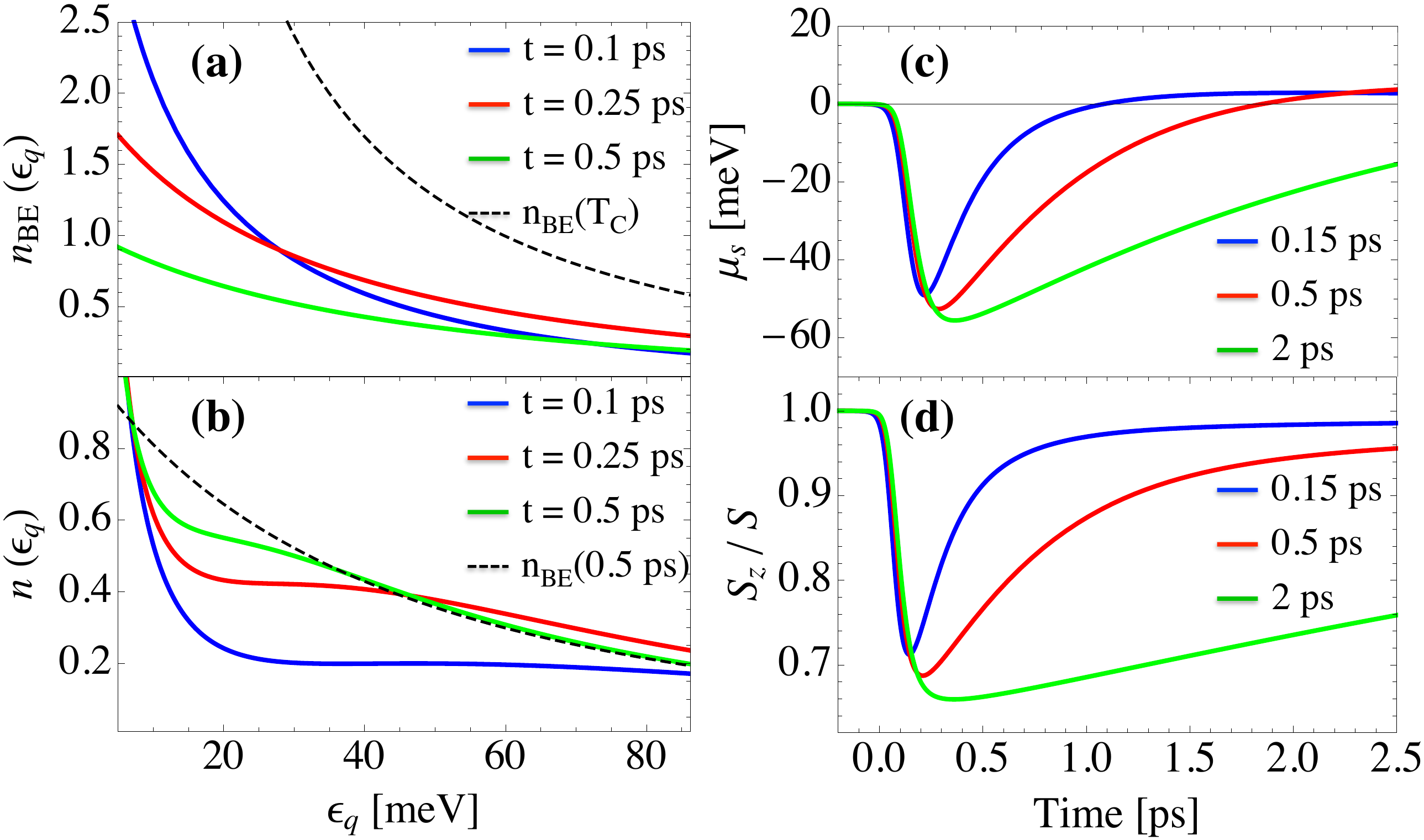}
\caption{Numerical solutions of Eqs.~(\ref{eq:museq}) and (\ref{eq:neq}) after the itinerant electrons are heated from $10^2\ \mathrm{K}$ to $10^3\ \mathrm{K}$ ($T_{C}$) within 50 fs. $\epsilon_{0}=5\ \mathrm{meV}$, $A=0.6\ \mathrm{meV\cdot nm^{2}}$, $\rho=6\ \mathrm{meV\cdot nm^{3}}$, $\tau_{s}=2\ \mathrm{ps}$, and $\alpha^*=10\alpha=0.1$. \textbf{(a)} The itinerant electron-hole pair distribution $n_{\mathrm{BE}}(\epsilon-\mu_{s})$ is rapidly depleted by a large spin accumulation $\mu_{s}$ that is built up via electron-magnon scattering. \textbf{(b)} The non-thermalized magnon distribution function $n(\epsilon_{q})$ equilibrates toward $n_{\mathrm{BE}}(\epsilon_{q}-\mu_{s})$. However, whereas the high-energy magnon states are rapidly populated, the low-energy states remain unaffected on short time scales. \textbf{(c)} Time evolution of the out-of-equilibrium spin accumulation, $\mu_{s}(t)$, and \textbf{(d)} the longitudinal spin density, $-S_{z}(t)$, after the temperature is exponentially relaxed toward $10^{2}\ \mathrm{K}$ on time scales of 0.15 ps, 0.5 ps, and 2 ps.}
\label{fig:distfunc}
\end{figure}

In the opposite high-frequency limit, pertinent to ultrafast demagnetization experiments, we consider F to be in a low-temperature equilibrium state before being excited by a THz laser pulse at $t=0$, upon which the effective temperature of the itinerant electron bath instantly increases such that $T_{\mathrm{eff}}\gtrsim T_{C}$. This regime is clearly beyond the validity of the LLG phenomenology, which is designed to address the low-energy extremum of magnetization dynamics. Dissipation in the LLG equation, including relaxation terms based on the stochastic Landau-Lifshitz-Bloch treatment~\cite{garaninPRB97, *PhysRevB.81.174401,mentinkPRL12}, is subject to a simple Markovian environment without any feedback or internal dynamics. This perspective must be refined for high frequencies when no subsystem can be viewed as a featureless reservoir for energy and angular momentum. Ultrafast heating of F rapidly excites the itinerant electron bath to a far-from-equilibrium state. The subsequent buildup of $\mu_{s}$ via electron-magnon scattering (\ref{eq:sdcurrent}) rapidly depletes the number of available scattering states; see Fig.~\ref{fig:distfunc}a. Consequently, the total relaxation of spin angular momentum from the combined electronic system is ultimately bounded by $\tau_{s}^{-1}$.

To appreciate the non-thermalized nature of the excited magnons, we consider the limit in which $\mu_{s}$ is small compared with the magnon gap $\epsilon_{0}$ and no magnons are excited [$n(\epsilon_{q})=0$] for $t<0$. After rapid heating of the itinerant electron bath at $t=0$, the time evolution of the non-thermalized magnon distribution follows
\begin{equation}
\label{eq:neqtime}
n(\epsilon_{q},t)\approx n_{BE}(\epsilon_{q},t)\left[1-e^{-\Gamma(\epsilon_{q})\epsilon_{q} t}\right].
\end{equation}
This result implies that high-energy states are populated much faster than are low-energy states; see Fig.~\ref{fig:distfunc}b for numerical solutions of Eqs. (\ref{eq:museq}) and (\ref{eq:neq}) when $T_{\mathrm{eff}}$ is increased from $10^2\ \mathrm{K}$ to $10^3\ \mathrm{K}$ within 50 fs and the effects of a finite $\mu_{s}$ are included. By comparison, internal magnon-magnon interactions equilibrate the distribution function on the time scale $\tau_{eq}^{-1}\sim \hbar^{-1}\epsilon_{m}[\epsilon_{m}/(k_{B}T_{C})]^{3}$~\cite{benderPRB14}, where $\epsilon_{m}$ is a characteristic energy of the thermal magnon cloud. For short times, $I_{sd}$ (\ref{eq:sdcurrent}) dominates the magnon dynamics, and we expect the magnon population to significantly differ from the thermalized Bose-Einstein distribution.

When $T_{\mathrm{eff}}>T_{C}$, the thermally excited electron-hole pairs are populated in accordance with the classical Rayleigh-Jeans distribution, $n_{BE}(\epsilon_{q}-\mu_{s})\to k_{B}T_{\mathrm{eff}}/(\epsilon_{q}-\mu_{s})$. Assuming, for simplicity, that the expansion for $\Gamma(\epsilon_{q})$ is valid throughout the Brillouin zone, Eq.~(\ref{eq:sdcurrent}) yields $\partial_{t}n_{d}|_{t\to0}=I_{sd}(0)/\hbar=[\Gamma_{0}+3\chi(\epsilon_{b}-\epsilon_{0})/5]k_{B}T_{\mathrm{eff}}S/\hbar$. Thus, the demagnetization rate of the $d$ orbitals is initially proportional to the temperature of the electron bath but is rapidly reduced by the lack of available scattering states for high-energy magnons within the time scale of the demagnetization process. This finding conflicts with the results obtained from a Langevin treatment of the LLG equation~\cite{brownPR63,*Kubo01061970}, in which the magnetization relaxation rate is proportional to the temperature difference at \textit{all times}. Figures \ref{fig:distfunc}c and \ref{fig:distfunc}d illustrate the time evolution of the out-of-equilibrium spin accumulation, $\mu_{s}(t)$, and the longitudinal spin density, $-S_{z}(t)$, for different relaxation times of $T_{\mathrm{eff}}$.

In the ultrafast regime, the electron-magnon spin-flop scattering is governed by the \textit{effective} Gilbert damping parameter $\alpha^{*}\equiv \chi (\epsilon_{b}-\epsilon_{0})$. Recent experimental investigations of the magnon relaxation rates on Co and Fe surfaces confirm that high-$q$ magnons have significantly shorter lifetimes than do low-$q$ magnons~\cite{zhangPRL12rt}. It is reasonable to assume that the same effects are also present in the bulk. The initial relaxation time scale in the ultrafast regime is $\tau_i \sim (\alpha^{*}\hbar^{-1}k_{B}T_{\mathrm{eff}})^{-1}$. This generalizes the result of Koopmans \textit{et al.}~\cite{koopmansPRL05} for the ultrafast relaxation of the longitudinal magnetization to arbitrary $\alpha^{*}$ based on the transverse spin diffusion~\cite{tserkovPRB09sd}. The notion of magnons becomes questionable when the intrinsic linewidth approaches the magnon energy, which corresponds to $\alpha^{*}\sim 1$. Staying well below this limit, and consistent with Refs.~\cite{zhangPRL12rt,tserkovPRB09sd}, we use $\alpha^{*}=0.1$. For $T_{C}= 10^3$ K the initial relaxation time scale $\tau_i \sim 10^{2}(T_{C}/T_{\mathrm{eff}})$ fs, which is generally consistent with the demagnetization rates observed for ultrafast demagnetization in Fe~\cite{PhysRevB.65.104429,*PhysRevB.78.174422,*PhysRevB.84.132412,*mathiasPNAS12}.

\begin{figure}
\centering
\includegraphics[scale=0.4]{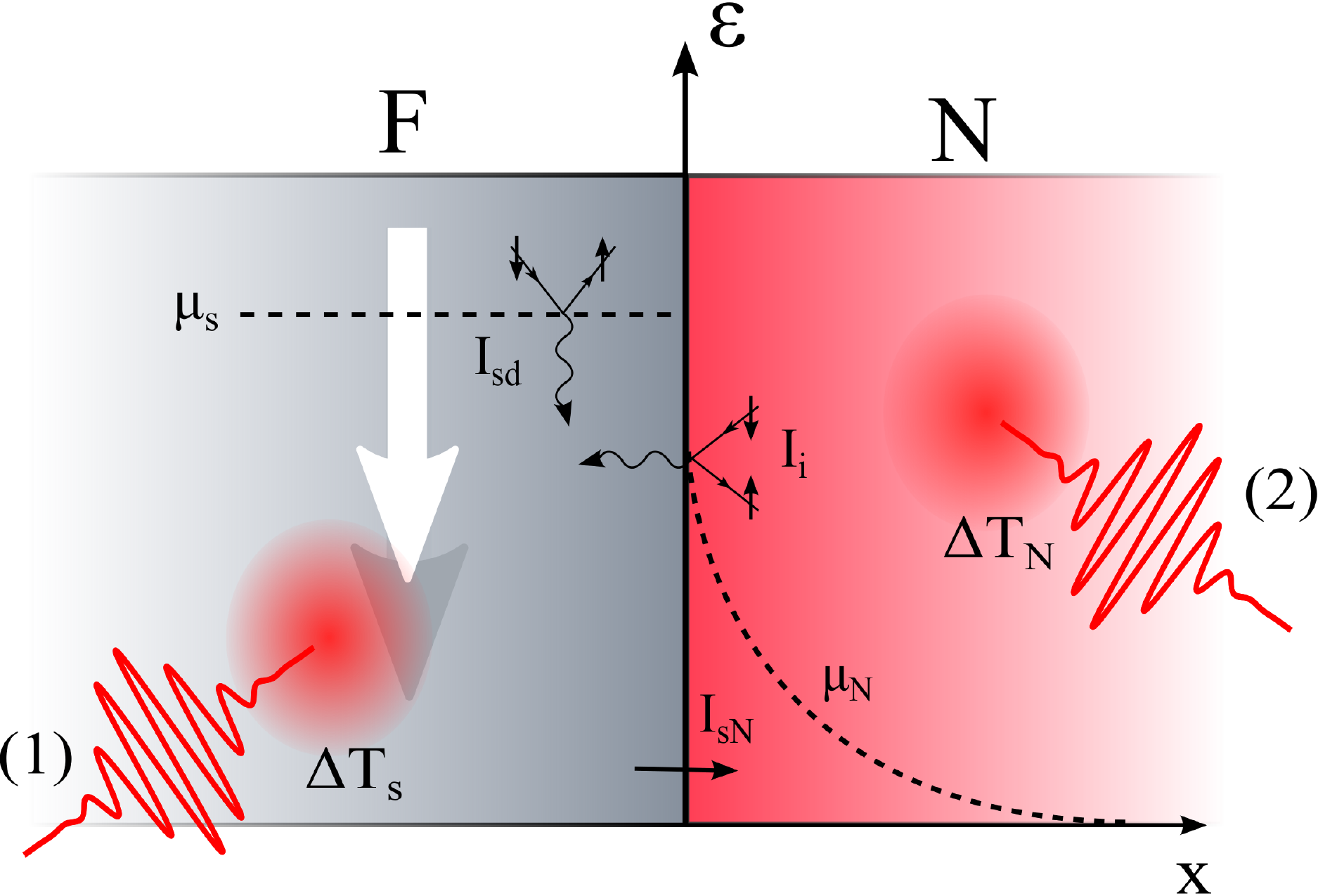}
\caption{Sketch of a metallic ferromagnet (F) coupled to a normal metal (N). In the ultrafast regime, both the rapid heating of $s$ electrons in F by $\Delta T_{s}$ (scenario (1)) and the heating of N by $\Delta T_{\mathrm{N}}$ (scenario (2)) demagnetize the $d$ electrons in F. $I_{sd}$ (\ref{eq:sdcurrent}) leads to the $s$ electron spin accumulation $\mu_{s}$ in F, whereas $I_{i}$ (\ref{eq:icurrent}) leads to the spin accumulation $\mu_{\mathrm{N}}^{0}$ at the F$|$N interface. Subsequently, $\mu_{\mathrm{N}}(x)$ diffuses into N until it vanishes due to spin-flip dissipation to the lattice. The additional interfacial spin current $I_{s\mathrm{N}}$ between the $s$ electrons in F and the itinerant electrons in N, due to the thermodynamic biases $\delta\mu=\mu_{s}-\mu_{\mathrm{N}}^{0}$ and $\delta T=T_{s}-T_{\mathrm{N}}$, can be described by conventional thermoelectric parameters for longitudinal spin-dependent transport~\cite{PhysRevLett.99.066603}.}
\label{fig:heterostructure}
\end{figure}

Figure \ref{fig:heterostructure} presents a schematic illustration of an F$|$N interface. In magnetic heterostructures, and for stand-alone ferromagnets on a conducting substrate, the demagnetization dynamics of F are also affected by the spin accumulation in N, $\mu_{\mathrm{N}}(x)$, which can impact how non-local laser irradiation (e.g.,~the heating of N alone) induces ultrafast demagnetization of F~\cite{eschenlohrNATM13}. By adding terms of the form $\sim\sum_{qkk'} U_{qkk'}a_{q}\tilde{c}_{k\uparrow}^{\dagger}\tilde{c}_{k'\downarrow}$ to $\hat{H}_{sd}$, where $\tilde{c}_{k\uparrow}^{\dagger}$ $(\tilde{c}_{k'\downarrow})$ describes the creation (annihilation) of an electron with spin up (down) at the F$|$N interface, the \textit{interfacial} spin transfer (per unit area) due to electron-magnon spin-flop scattering is~\cite{benderPRL12}
\begin{equation}
\label{eq:icurrent}
I_{i}=\int_{\epsilon_0}^{\epsilon_{b}}d\epsilon_{q} \Gamma^{i} (\epsilon_{q}) (\epsilon_{q}-\mu^{0}_{\mathrm{N}})\mathcal{D}(\epsilon_{q})\left[n_{\rm BE}(\epsilon_{q}-\mu^{0}_{\mathrm{N}})-n(\epsilon_{q})\right],
\end{equation}
where $\mu_{\mathrm{N}}^{0}\equiv \mu_{\mathrm{N}}(0)$ is the spin accumulation at the interface and $\Gamma^{i}(\epsilon_{q})$ parameterizes the interfacial scattering rate. 

Importantly, we note that the scattering of coherent long-wavelength magnons at the F$|$N interface can be described in the language of spin pumping/spin Seebeck effects~\cite{benderPRL12, benderPRB14} and can be parameterized in terms of the spin-mixing conductance $g_{\uparrow \downarrow}$ (per unit area)~\cite{brataasPRL00}. Motivated by $\Gamma(\epsilon_{q})$ in the bulk, we write $\Gamma^{i}(\epsilon_{q})=g_{\uparrow \downarrow}^{*}(\epsilon_{q})/(\pi S)$, where $g_{\uparrow \downarrow}^{*}$ reduces to $g_{\uparrow \downarrow}$ for low-energy scattering, $\epsilon_{q}\to 0$. The interface scattering (\ref{eq:icurrent}) dominates the microwave spin relaxation in thin ferromagnetic layers of thickness $d_{\mathrm{F}}\lesssim 10\ \mathrm{nm}$~\cite{mizukamiJJAP01,*mizukamiPRB02,*PhysRevLett.113.207202,tserkovPRL02sp}. This trend should continue for higher frequencies and is relevant for ultrafast spin dynamics in thin magnetic layers in heterostructures~\cite{melnikovPRL11,*turgutPRL13,*choiNC14}. We expect the energy dependence of the effective interfacial spin-mixing conductance to be relatively weak compared to that of the bulk scattering $\Gamma(\epsilon_{q})$, which can be severely constrained at low energies due to momentum conservation~\cite{tserkovPRB09sd}. For a finite temperature bias $\delta T$ across the interface and for magnons thermalized at the temperature $T<T_{C}$, the connection to the thermal spin Seebeck and Peltier effects is made by identifying $\mathcal{S}=\partial_{T} I_{i}$ and $\Pi=T\mathcal{S}/\hbar$~\cite{hoffmanPRB13} as the Seebeck and Peltier coefficients, respectively. In the ultrafast regime, the relative importance of the bulk scattering, parameterized by $\alpha^{*}$, and the interfacial scattering, parameterized by $g_{\uparrow \downarrow}^{*}$, can be extracted from measurements of demagnetization strength and spin currents in magnetic heterostructures. 

In conclusion, we have extended the concepts of transverse spin diffusion in bulk ferromagnets and the spin-mixing physics across interfaces to address the ultrafast spin dynamics observed in rapidly heated magnetic heterostructures. In the microwave, thermal, and ultrafast regimes the elementary interactions that defines the demagnetization time scale are described by kinetic rate equations for electron-magnon spin-flop scattering. For metallic ferromagnets in the bulk, our analysis shows that treating the itinerant and localized electron subsystems as being individually thermalized with equilibrium distribution functions parameterized by an effective temperature is insufficient to describe the far-from-equilibrium spin dynamics that arise from pulsed laser heating. The magnon distribution function remains non-thermalized on the relevant time scale of the demagnetization process. We emphasize the critical role of the out-of-equilibrium spin accumulation $\mu_{s}$ among the itinerant electrons, which provides the bottleneck that limits the relaxation of spin angular momentum from the combined electronic system.

The authors thank A.~V.~Kimel, G.~E.~W.~Bauer, J.~Barker, S.~Bender, and H.~Skarsv\aa g for valuable discussions.


\begin{thebibliography}{52}%
\makeatletter
\providecommand \@ifxundefined [1]{%
 \@ifx{#1\undefined}
}%
\providecommand \@ifnum [1]{%
 \ifnum #1\expandafter \@firstoftwo
 \else \expandafter \@secondoftwo
 \fi
}%
\providecommand \@ifx [1]{%
 \ifx #1\expandafter \@firstoftwo
 \else \expandafter \@secondoftwo
 \fi
}%
\providecommand \natexlab [1]{#1}%
\providecommand \enquote  [1]{``#1''}%
\providecommand \bibnamefont  [1]{#1}%
\providecommand \bibfnamefont [1]{#1}%
\providecommand \citenamefont [1]{#1}%
\providecommand \href@noop [0]{\@secondoftwo}%
\providecommand \href [0]{\begingroup \@sanitize@url \@href}%
\providecommand \@href[1]{\@@startlink{#1}\@@href}%
\providecommand \@@href[1]{\endgroup#1\@@endlink}%
\providecommand \@sanitize@url [0]{\catcode `\\12\catcode `\$12\catcode
  `\&12\catcode `\#12\catcode `\^12\catcode `\_12\catcode `\%12\relax}%
\providecommand \@@startlink[1]{}%
\providecommand \@@endlink[0]{}%
\providecommand \url  [0]{\begingroup\@sanitize@url \@url }%
\providecommand \@url [1]{\endgroup\@href {#1}{\urlprefix }}%
\providecommand \urlprefix  [0]{URL }%
\providecommand \Eprint [0]{\href }%
\@ifxundefined \urlstyle {%
  \providecommand \doi  [0]{\begingroup \@sanitize@url \@doi}%
  \providecommand \@doi [1]{\endgroup \@@startlink {\doibase
  #1}doi:\discretionary {}{}{}#1\@@endlink }%
}{%
  \providecommand \doi  [0]{doi:\discretionary{}{}{}\begingroup
  \urlstyle{rm}\Url }%
}%
\providecommand \doibase [0]{http://dx.doi.org/}%
\providecommand \Doi [0]{\begingroup \@sanitize@url \@Doi }%
\providecommand \@Doi  [1]{\endgroup\@@startlink{\doibase#1}\@@Doi}%
\providecommand \@@Doi [1]{#1\@@endlink}%
\providecommand \selectlanguage [0]{\@gobble}%
\providecommand \bibinfo  [0]{\@secondoftwo}%
\providecommand \bibfield  [0]{\@secondoftwo}%
\providecommand \translation [1]{[#1]}%
\providecommand \BibitemOpen [0]{}%
\providecommand \bibitemStop [0]{}%
\providecommand \bibitemNoStop [0]{.\EOS\space}%
\providecommand \EOS [0]{\spacefactor3000\relax}%
\providecommand \BibitemShut  [1]{\csname bibitem#1\endcsname}%
\bibitem [{\citenamefont {Melnikov}\ \emph {et~al.}(2011)\citenamefont
  {Melnikov}, \citenamefont {Razdolski}, \citenamefont {Wehling}, \citenamefont
  {Papaioannou}, \citenamefont {Roddatis}, \citenamefont {Fumagalli},
  \citenamefont {Aktsipetrov}, \citenamefont {Lichtenstein},\ and\
  \citenamefont {Bovensiepen}}]{melnikovPRL11}%
  \BibitemOpen
  \bibfield  {author} {\bibinfo {author} {\bibfnamefont {A.}~\bibnamefont
  {Melnikov}}, \bibinfo {author} {\bibfnamefont {I.}~\bibnamefont {Razdolski}},
  \bibinfo {author} {\bibfnamefont {T.~O.}\ \bibnamefont {Wehling}}, \bibinfo
  {author} {\bibfnamefont {E.~T.}\ \bibnamefont {Papaioannou}}, \bibinfo
  {author} {\bibfnamefont {V.}~\bibnamefont {Roddatis}}, \bibinfo {author}
  {\bibfnamefont {P.}~\bibnamefont {Fumagalli}}, \bibinfo {author}
  {\bibfnamefont {O.}~\bibnamefont {Aktsipetrov}}, \bibinfo {author}
  {\bibfnamefont {A.~I.}\ \bibnamefont {Lichtenstein}}, \ and\ \bibinfo
  {author} {\bibfnamefont {U.}~\bibnamefont {Bovensiepen}},\ }\Doi
  {10.1103/PhysRevLett.107.076601} {\bibfield  {journal} {\bibinfo  {journal}
  {Phys. Rev. Lett.},\ }\textbf {\bibinfo {volume} {107}},\ \bibinfo {pages}
  {076601} (\bibinfo {year} {2011})}\BibitemShut {NoStop}%
\bibitem [{\citenamefont {Turgut}\ \emph {et~al.}(2013)\citenamefont {Turgut},
  \citenamefont {o~vorakiat}, \citenamefont {Shaw}, \citenamefont {Grychtol},
  \citenamefont {Nembach}, \citenamefont {Rudolf}, \citenamefont {Adam},
  \citenamefont {Aeschlimann}, \citenamefont {Schneider}, \citenamefont
  {Silva}, \citenamefont {Murnane}, \citenamefont {Kapteyn},\ and\
  \citenamefont {Mathias}}]{turgutPRL13}%
  \BibitemOpen
  \bibfield  {author} {\bibinfo {author} {\bibfnamefont {E.}~\bibnamefont
  {Turgut}}, \bibinfo {author} {\bibfnamefont {C.~L.}\ \bibnamefont
  {o~vorakiat}}, \bibinfo {author} {\bibfnamefont {J.~M.}\ \bibnamefont
  {Shaw}}, \bibinfo {author} {\bibfnamefont {P.}~\bibnamefont {Grychtol}},
  \bibinfo {author} {\bibfnamefont {H.~T.}\ \bibnamefont {Nembach}}, \bibinfo
  {author} {\bibfnamefont {D.}~\bibnamefont {Rudolf}}, \bibinfo {author}
  {\bibfnamefont {R.}~\bibnamefont {Adam}}, \bibinfo {author} {\bibfnamefont
  {M.}~\bibnamefont {Aeschlimann}}, \bibinfo {author} {\bibfnamefont {C.~M.}\
  \bibnamefont {Schneider}}, \bibinfo {author} {\bibfnamefont {T.~J.}\
  \bibnamefont {Silva}}, \bibinfo {author} {\bibfnamefont {M.~M.}\ \bibnamefont
  {Murnane}}, \bibinfo {author} {\bibfnamefont {H.~C.}\ \bibnamefont
  {Kapteyn}}, \ and\ \bibinfo {author} {\bibfnamefont {S.}~\bibnamefont
  {Mathias}},\ }\Doi {10.1103/PhysRevLett.110.197201} {\bibfield  {journal}
  {\bibinfo  {journal} {Phys. Rev. Lett.},\ }\textbf {\bibinfo {volume}
  {110}},\ \bibinfo {pages} {197201} (\bibinfo {year} {2013})}\BibitemShut
  {NoStop}%
\bibitem [{\citenamefont {Choi}\ \emph {et~al.}(2014)\citenamefont {Choi},
  \citenamefont {Min}, \citenamefont {Lee},\ and\ \citenamefont
  {Cahill}}]{choiNC14}%
  \BibitemOpen
  \bibfield  {author} {\bibinfo {author} {\bibfnamefont {G.-M.}\ \bibnamefont
  {Choi}}, \bibinfo {author} {\bibfnamefont {B.-C.}\ \bibnamefont {Min}},
  \bibinfo {author} {\bibfnamefont {K.-J.}\ \bibnamefont {Lee}}, \ and\
  \bibinfo {author} {\bibfnamefont {D.~G.}\ \bibnamefont {Cahill}},\
  }\href@noop {} {\bibfield  {journal} {\bibinfo  {journal} {Nature Comm.},\
  }\textbf {\bibinfo {volume} {5}},\ \bibinfo {pages} {4334} (\bibinfo {year}
  {2014})}\BibitemShut {NoStop}%
\bibitem [{\citenamefont {Kirilyuk}\ \emph
  {et~al.}(2010){\natexlab{a}}\citenamefont {Kirilyuk}, \citenamefont {Kimel},\
  and\ \citenamefont {Rasing}}]{RevModPhys.82.2731}%
  \BibitemOpen
  \bibfield  {author} {\bibinfo {author} {\bibfnamefont {A.}~\bibnamefont
  {Kirilyuk}}, \bibinfo {author} {\bibfnamefont {A.~V.}\ \bibnamefont {Kimel}},
  \ and\ \bibinfo {author} {\bibfnamefont {T.}~\bibnamefont {Rasing}},\ }\Doi
  {10.1103/RevModPhys.82.2731} {\bibfield  {journal} {\bibinfo  {journal} {Rev.
  Mod. Phys.},\ }\textbf {\bibinfo {volume} {82}},\ \bibinfo {pages} {2731}
  (\bibinfo {year} {2010}{\natexlab{a}})}\BibitemShut {NoStop}%
\bibitem [{\citenamefont {Lifshitz}\ and\ \citenamefont
  {Pitaevskii}(1980)}]{landauBOOKv9}%
  \BibitemOpen
  \bibfield  {author} {\bibinfo {author} {\bibfnamefont {E.~M.}\ \bibnamefont
  {Lifshitz}}\ and\ \bibinfo {author} {\bibfnamefont {L.~P.}\ \bibnamefont
  {Pitaevskii}},\ }\href@noop {} {\emph {\bibinfo {title} {Statistical Physics,
  Part 2}}},\ \bibinfo {edition} {3rd}\ ed.,\ \bibinfo {series} {Course of
  Theoretical Physics}, Vol.~\bibinfo {volume} {9}\ (\bibinfo  {publisher}
  {Pergamon},\ \bibinfo {address} {Oxford},\ \bibinfo {year}
  {1980})\BibitemShut {NoStop}%
\bibitem [{\citenamefont {Gilbert}(2004)}]{gilbertIEEEM04}%
  \BibitemOpen
  \bibfield  {author} {\bibinfo {author} {\bibfnamefont {T.~L.}\ \bibnamefont
  {Gilbert}},\ }\href@noop {} {\bibfield  {journal} {\bibinfo  {journal} {IEEE
  Trans. Magn.},\ }\textbf {\bibinfo {volume} {40}},\ \bibinfo {pages} {3443}
  (\bibinfo {year} {2004})}\BibitemShut {NoStop}%
\bibitem [{\citenamefont {Kittel}(1948)}]{PhysRev.73.155}%
  \BibitemOpen
  \bibfield  {author} {\bibinfo {author} {\bibfnamefont {C.}~\bibnamefont
  {Kittel}},\ }\Doi {10.1103/PhysRev.73.155} {\bibfield  {journal} {\bibinfo
  {journal} {Phys. Rev.},\ }\textbf {\bibinfo {volume} {73}},\ \bibinfo {pages}
  {155} (\bibinfo {year} {1948})}\BibitemShut {NoStop}%
\bibitem [{\citenamefont {Uchida}\ \emph {et~al.}(2010)\citenamefont {Uchida},
  \citenamefont {Xiao}, \citenamefont {Adachi}, \citenamefont {Ohe},
  \citenamefont {Takahashi}, \citenamefont {Ieda}, \citenamefont {Ota},
  \citenamefont {Kajiwara}, \citenamefont {Umezawa}, \citenamefont {Kawai},
  \citenamefont {Bauer}, \citenamefont {Maekawa},\ and\ \citenamefont
  {Saitoh}}]{uchidaNATM10}%
  \BibitemOpen
  \bibfield  {author} {\bibinfo {author} {\bibfnamefont {K.}~\bibnamefont
  {Uchida}}, \bibinfo {author} {\bibfnamefont {J.}~\bibnamefont {Xiao}},
  \bibinfo {author} {\bibfnamefont {H.}~\bibnamefont {Adachi}}, \bibinfo
  {author} {\bibfnamefont {J.}~\bibnamefont {Ohe}}, \bibinfo {author}
  {\bibfnamefont {S.}~\bibnamefont {Takahashi}}, \bibinfo {author}
  {\bibfnamefont {J.}~\bibnamefont {Ieda}}, \bibinfo {author} {\bibfnamefont
  {T.}~\bibnamefont {Ota}}, \bibinfo {author} {\bibfnamefont {Y.}~\bibnamefont
  {Kajiwara}}, \bibinfo {author} {\bibfnamefont {H.}~\bibnamefont {Umezawa}},
  \bibinfo {author} {\bibfnamefont {H.}~\bibnamefont {Kawai}}, \bibinfo
  {author} {\bibfnamefont {G.~E.~W.}\ \bibnamefont {Bauer}}, \bibinfo {author}
  {\bibfnamefont {S.}~\bibnamefont {Maekawa}}, \ and\ \bibinfo {author}
  {\bibfnamefont {E.}~\bibnamefont {Saitoh}},\ }\href@noop {} {\bibfield
  {journal} {\bibinfo  {journal} {Nat. Mater.},\ }\textbf {\bibinfo {volume}
  {9}},\ \bibinfo {pages} {894} (\bibinfo {year} {2010})}\BibitemShut {NoStop}%
\bibitem [{\citenamefont {Xiao}\ \emph {et~al.}(2010)\citenamefont {Xiao},
  \citenamefont {Bauer}, \citenamefont {Uchida}, \citenamefont {Saitoh},\ and\
  \citenamefont {Maekawa}}]{xiaoPRB10}%
  \BibitemOpen
  \bibfield  {author} {\bibinfo {author} {\bibfnamefont {J.}~\bibnamefont
  {Xiao}}, \bibinfo {author} {\bibfnamefont {G.~E.~W.}\ \bibnamefont {Bauer}},
  \bibinfo {author} {\bibfnamefont {K.}~\bibnamefont {Uchida}}, \bibinfo
  {author} {\bibfnamefont {E.}~\bibnamefont {Saitoh}}, \ and\ \bibinfo {author}
  {\bibfnamefont {S.}~\bibnamefont {Maekawa}},\ }\Doi
  {10.1103/PhysRevB.81.214418} {\bibfield  {journal} {\bibinfo  {journal}
  {Phys. Rev. B},\ }\textbf {\bibinfo {volume} {81}},\ \bibinfo {pages}
  {214418} (\bibinfo {year} {2010})}\BibitemShut {NoStop}%
\bibitem [{\citenamefont {Bauer}\ \emph {et~al.}(2012)\citenamefont {Bauer},
  \citenamefont {Saitoh},\ and\ \citenamefont {{van Wees}}}]{bauerNATM12}%
  \BibitemOpen
  \bibfield  {author} {\bibinfo {author} {\bibfnamefont {G.~E.~W.}\
  \bibnamefont {Bauer}}, \bibinfo {author} {\bibfnamefont {E.}~\bibnamefont
  {Saitoh}}, \ and\ \bibinfo {author} {\bibfnamefont {B.~J.}\ \bibnamefont
  {{van Wees}}},\ }\href@noop {} {\bibfield  {journal} {\bibinfo  {journal}
  {Nat. Mater.},\ }\textbf {\bibinfo {volume} {11}},\ \bibinfo {pages} {391}
  (\bibinfo {year} {2012})}\BibitemShut {NoStop}%
\bibitem [{\citenamefont {Koopmans}\ \emph {et~al.}(2005)\citenamefont
  {Koopmans}, \citenamefont {Ruigrok}, \citenamefont {{Dalla Longa}},\ and\
  \citenamefont {de~Jonge}}]{koopmansPRL05}%
  \BibitemOpen
  \bibfield  {author} {\bibinfo {author} {\bibfnamefont {B.}~\bibnamefont
  {Koopmans}}, \bibinfo {author} {\bibfnamefont {J.~J.~M.}\ \bibnamefont
  {Ruigrok}}, \bibinfo {author} {\bibfnamefont {F.}~\bibnamefont {{Dalla
  Longa}}}, \ and\ \bibinfo {author} {\bibfnamefont {W.~J.~M.}\ \bibnamefont
  {de~Jonge}},\ }\href@noop {} {\bibfield  {journal} {\bibinfo  {journal}
  {Phys. Rev. Lett.},\ }\textbf {\bibinfo {volume} {95}},\ \bibinfo {eid}
  {267207} (\bibinfo {year} {2005})}\BibitemShut {NoStop}%
\bibitem [{\citenamefont {Walowski}\ \emph {et~al.}(2008)\citenamefont
  {Walowski}, \citenamefont {M\"uller}, \citenamefont {Djordjevic},
  \citenamefont {M\"unzenberg}, \citenamefont {Kl\"aui}, \citenamefont {Vaz},\
  and\ \citenamefont {Bland}}]{PhysRevLett.101.237401}%
  \BibitemOpen
  \bibfield  {author} {\bibinfo {author} {\bibfnamefont {J.}~\bibnamefont
  {Walowski}}, \bibinfo {author} {\bibfnamefont {G.}~\bibnamefont {M\"uller}},
  \bibinfo {author} {\bibfnamefont {M.}~\bibnamefont {Djordjevic}}, \bibinfo
  {author} {\bibfnamefont {M.}~\bibnamefont {M\"unzenberg}}, \bibinfo {author}
  {\bibfnamefont {M.}~\bibnamefont {Kl\"aui}}, \bibinfo {author} {\bibfnamefont
  {C.~A.~F.}\ \bibnamefont {Vaz}}, \ and\ \bibinfo {author} {\bibfnamefont
  {J.~A.~C.}\ \bibnamefont {Bland}},\ }\Doi {10.1103/PhysRevLett.101.237401}
  {\bibfield  {journal} {\bibinfo  {journal} {Phys. Rev. Lett.},\ }\textbf
  {\bibinfo {volume} {101}},\ \bibinfo {pages} {237401} (\bibinfo {year}
  {2008})}\BibitemShut {NoStop}%
\bibitem [{\citenamefont {Brataas}\ \emph {et~al.}(2012)\citenamefont
  {Brataas}, \citenamefont {Kent},\ and\ \citenamefont
  {Ohno}}]{Brataas:2012fk}%
  \BibitemOpen
  \bibfield  {author} {\bibinfo {author} {\bibfnamefont {A.}~\bibnamefont
  {Brataas}}, \bibinfo {author} {\bibfnamefont {A.~D.}\ \bibnamefont {Kent}}, \
  and\ \bibinfo {author} {\bibfnamefont {H.}~\bibnamefont {Ohno}},\ }\href
  {http://dx.doi.org/10.1038/nmat3311} {\bibfield  {journal} {\bibinfo
  {journal} {Nature Mater.},\ }\textbf {\bibinfo {volume} {11}},\ \bibinfo
  {pages} {372} (\bibinfo {year} {2012})}\BibitemShut {NoStop}%
\bibitem [{\citenamefont {Beaurepaire}\ \emph {et~al.}(1996)\citenamefont
  {Beaurepaire}, \citenamefont {Merle}, \citenamefont {Daunois},\ and\
  \citenamefont {Bigot}}]{beaurepairePRL96}%
  \BibitemOpen
  \bibfield  {author} {\bibinfo {author} {\bibfnamefont {E.}~\bibnamefont
  {Beaurepaire}}, \bibinfo {author} {\bibfnamefont {J.-C.}\ \bibnamefont
  {Merle}}, \bibinfo {author} {\bibfnamefont {A.}~\bibnamefont {Daunois}}, \
  and\ \bibinfo {author} {\bibfnamefont {J.-Y.}\ \bibnamefont {Bigot}},\ }\Doi
  {10.1103/PhysRevLett.76.4250} {\bibfield  {journal} {\bibinfo  {journal}
  {Phys. Rev. Lett.},\ }\textbf {\bibinfo {volume} {76}},\ \bibinfo {pages}
  {4250} (\bibinfo {year} {1996})}\BibitemShut {NoStop}%
\bibitem [{\citenamefont {{van der Ziel}}\ \emph {et~al.}(1965)\citenamefont
  {{van der Ziel}}, \citenamefont {Pershan},\ and\ \citenamefont
  {Malmstrom}}]{zielPRL65}%
  \BibitemOpen
  \bibfield  {author} {\bibinfo {author} {\bibfnamefont {J.~P.}\ \bibnamefont
  {{van der Ziel}}}, \bibinfo {author} {\bibfnamefont {P.~S.}\ \bibnamefont
  {Pershan}}, \ and\ \bibinfo {author} {\bibfnamefont {L.~D.}\ \bibnamefont
  {Malmstrom}},\ }\Doi {10.1103/PhysRevLett.15.190} {\bibfield  {journal}
  {\bibinfo  {journal} {Phys. Rev. Lett.},\ }\textbf {\bibinfo {volume} {15}},\
  \bibinfo {pages} {190} (\bibinfo {year} {1965})}\BibitemShut {NoStop}%
\bibitem [{\citenamefont {Zhang}\ and\ \citenamefont
  {H{\"u}bner}(2000)}]{zhangPRL00}%
  \BibitemOpen
  \bibfield  {author} {\bibinfo {author} {\bibfnamefont {G.~P.}\ \bibnamefont
  {Zhang}}\ and\ \bibinfo {author} {\bibfnamefont {W.}~\bibnamefont
  {H{\"u}bner}},\ }\Doi {10.1103/PhysRevLett.85.3025} {\bibfield  {journal}
  {\bibinfo  {journal} {Phys. Rev. Lett.},\ }\textbf {\bibinfo {volume} {85}},\
  \bibinfo {pages} {3025} (\bibinfo {year} {2000})}\BibitemShut {NoStop}%
\bibitem [{\citenamefont {Bigot}\ \emph {et~al.}(2009)\citenamefont {Bigot},
  \citenamefont {Vomir},\ and\ \citenamefont {Beaurepaire}}]{bigotNATP09}%
  \BibitemOpen
  \bibfield  {author} {\bibinfo {author} {\bibfnamefont {J.-Y.}\ \bibnamefont
  {Bigot}}, \bibinfo {author} {\bibfnamefont {M.}~\bibnamefont {Vomir}}, \ and\
  \bibinfo {author} {\bibfnamefont {E.}~\bibnamefont {Beaurepaire}},\
  }\href@noop {} {\bibfield  {journal} {\bibinfo  {journal} {Nature Phys.},\
  }\textbf {\bibinfo {volume} {5}},\ \bibinfo {pages} {515} (\bibinfo {year}
  {2009})}\BibitemShut {NoStop}%
\bibitem [{\citenamefont {Kirilyuk}\ \emph
  {et~al.}(2010){\natexlab{b}}\citenamefont {Kirilyuk}, \citenamefont {Kimel},\
  and\ \citenamefont {{Th. Rasing}}}]{kirilyukRMP10}%
  \BibitemOpen
  \bibfield  {author} {\bibinfo {author} {\bibfnamefont {A.}~\bibnamefont
  {Kirilyuk}}, \bibinfo {author} {\bibfnamefont {A.~V.}\ \bibnamefont {Kimel}},
  \ and\ \bibinfo {author} {\bibnamefont {{Th. Rasing}}},\ }\Doi
  {10.1103/RevModPhys.82.2731} {\bibfield  {journal} {\bibinfo  {journal} {Rev.
  Mod. Phys.},\ }\textbf {\bibinfo {volume} {82}},\ \bibinfo {pages} {2731}
  (\bibinfo {year} {2010}{\natexlab{b}})}\BibitemShut {NoStop}%
\bibitem [{\citenamefont {Koopmans}\ \emph {et~al.}(2010)\citenamefont
  {Koopmans}, \citenamefont {Malinowski}, \citenamefont {{Dalla Longa}},
  \citenamefont {Steiauf}, \citenamefont {F{\"a}hnle}, \citenamefont {Roth},
  \citenamefont {Cinchetti},\ and\ \citenamefont
  {Aeschlimann}}]{koopmansNATM10}%
  \BibitemOpen
  \bibfield  {author} {\bibinfo {author} {\bibfnamefont {B.}~\bibnamefont
  {Koopmans}}, \bibinfo {author} {\bibfnamefont {G.}~\bibnamefont
  {Malinowski}}, \bibinfo {author} {\bibfnamefont {F.}~\bibnamefont {{Dalla
  Longa}}}, \bibinfo {author} {\bibfnamefont {D.}~\bibnamefont {Steiauf}},
  \bibinfo {author} {\bibfnamefont {M.}~\bibnamefont {F{\"a}hnle}}, \bibinfo
  {author} {\bibfnamefont {T.}~\bibnamefont {Roth}}, \bibinfo {author}
  {\bibfnamefont {M.}~\bibnamefont {Cinchetti}}, \ and\ \bibinfo {author}
  {\bibfnamefont {M.}~\bibnamefont {Aeschlimann}},\ }\href@noop {} {\bibfield
  {journal} {\bibinfo  {journal} {Nat. Mater.},\ }\textbf {\bibinfo {volume}
  {9}},\ \bibinfo {pages} {259} (\bibinfo {year} {2010})}\BibitemShut {NoStop}%
\bibitem [{\citenamefont {Schellekens}\ and\ \citenamefont
  {Koopmans}(2013)}]{schellekensPRB13}%
  \BibitemOpen
  \bibfield  {author} {\bibinfo {author} {\bibfnamefont {A.~J.}\ \bibnamefont
  {Schellekens}}\ and\ \bibinfo {author} {\bibfnamefont {B.}~\bibnamefont
  {Koopmans}},\ }\Doi {10.1103/PhysRevB.87.020407} {\bibfield  {journal}
  {\bibinfo  {journal} {Phys. Rev. B},\ }\textbf {\bibinfo {volume} {87}},\
  \bibinfo {pages} {020407} (\bibinfo {year} {2013})}\BibitemShut {NoStop}%
\bibitem [{\citenamefont {Mentink}\ \emph {et~al.}(2012)\citenamefont
  {Mentink}, \citenamefont {Hellsvik}, \citenamefont {Afanasiev}, \citenamefont
  {Ivanov}, \citenamefont {Kirilyuk}, \citenamefont {Kimel}, \citenamefont
  {Eriksson}, \citenamefont {Katsnelson},\ and\ \citenamefont {{Th.
  Rasing}}}]{mentinkPRL12}%
  \BibitemOpen
  \bibfield  {author} {\bibinfo {author} {\bibfnamefont {J.~H.}\ \bibnamefont
  {Mentink}}, \bibinfo {author} {\bibfnamefont {J.}~\bibnamefont {Hellsvik}},
  \bibinfo {author} {\bibfnamefont {D.~V.}\ \bibnamefont {Afanasiev}}, \bibinfo
  {author} {\bibfnamefont {B.~A.}\ \bibnamefont {Ivanov}}, \bibinfo {author}
  {\bibfnamefont {A.}~\bibnamefont {Kirilyuk}}, \bibinfo {author}
  {\bibfnamefont {A.~V.}\ \bibnamefont {Kimel}}, \bibinfo {author}
  {\bibfnamefont {O.}~\bibnamefont {Eriksson}}, \bibinfo {author}
  {\bibfnamefont {M.~I.}\ \bibnamefont {Katsnelson}}, \ and\ \bibinfo {author}
  {\bibnamefont {{Th. Rasing}}},\ }\Doi {10.1103/PhysRevLett.108.057202}
  {\bibfield  {journal} {\bibinfo  {journal} {Phys. Rev. Lett.},\ }\textbf
  {\bibinfo {volume} {108}},\ \bibinfo {pages} {057202} (\bibinfo {year}
  {2012})}\BibitemShut {NoStop}%
\bibitem [{\citenamefont {Eschenlohr}\ \emph {et~al.}(2013)\citenamefont
  {Eschenlohr}, \citenamefont {Battiato}, \citenamefont {Maldonado},
  \citenamefont {Pontius}, \citenamefont {Kachel}, \citenamefont {Holldack},
  \citenamefont {Mitzner}, \citenamefont {F{\"o}hlisch}, \citenamefont
  {Oppeneer},\ and\ \citenamefont {Stamm}}]{eschenlohrNATM13}%
  \BibitemOpen
  \bibfield  {author} {\bibinfo {author} {\bibfnamefont {A.}~\bibnamefont
  {Eschenlohr}}, \bibinfo {author} {\bibfnamefont {M.}~\bibnamefont
  {Battiato}}, \bibinfo {author} {\bibfnamefont {P.}~\bibnamefont {Maldonado}},
  \bibinfo {author} {\bibfnamefont {N.}~\bibnamefont {Pontius}}, \bibinfo
  {author} {\bibfnamefont {T.}~\bibnamefont {Kachel}}, \bibinfo {author}
  {\bibfnamefont {K.}~\bibnamefont {Holldack}}, \bibinfo {author}
  {\bibfnamefont {R.}~\bibnamefont {Mitzner}}, \bibinfo {author} {\bibfnamefont
  {A.}~\bibnamefont {F{\"o}hlisch}}, \bibinfo {author} {\bibfnamefont {P.~M.}\
  \bibnamefont {Oppeneer}}, \ and\ \bibinfo {author} {\bibfnamefont
  {C.}~\bibnamefont {Stamm}},\ }\href@noop {} {\bibfield  {journal} {\bibinfo
  {journal} {Nat. Mater.},\ }\textbf {\bibinfo {volume} {12}},\ \bibinfo
  {pages} {332} (\bibinfo {year} {2013})}\BibitemShut {NoStop}%
\bibitem [{\citenamefont {Ostler}\ \emph {et~al.}(2012)\citenamefont {Ostler},
  \citenamefont {Barker}, \citenamefont {Evans}, \citenamefont {Chantrell},
  \citenamefont {Atxitia}, \citenamefont {Chubykalo-Fesenko}, \citenamefont
  {El~Moussaoui}, \citenamefont {Le~Guyader}, \citenamefont {Mengotti},
  \citenamefont {Heyderman}, \citenamefont {Nolting}, \citenamefont
  {Tsukamoto}, \citenamefont {Itoh}, \citenamefont {Afanasiev}, \citenamefont
  {Ivanov}, \citenamefont {Kalashnikova}, \citenamefont {Vahaplar},
  \citenamefont {Mentink}, \citenamefont {Kirilyuk}, \citenamefont {Rasing},\
  and\ \citenamefont {Kimel}}]{Ostler:2012sf}%
  \BibitemOpen
  \bibfield  {author} {\bibinfo {author} {\bibfnamefont {T.~A.}\ \bibnamefont
  {Ostler}}, \bibinfo {author} {\bibfnamefont {J.}~\bibnamefont {Barker}},
  \bibinfo {author} {\bibfnamefont {R.~F.~L.}\ \bibnamefont {Evans}}, \bibinfo
  {author} {\bibfnamefont {R.~W.}\ \bibnamefont {Chantrell}}, \bibinfo {author}
  {\bibfnamefont {U.}~\bibnamefont {Atxitia}}, \bibinfo {author} {\bibfnamefont
  {O.}~\bibnamefont {Chubykalo-Fesenko}}, \bibinfo {author} {\bibfnamefont
  {S.}~\bibnamefont {El~Moussaoui}}, \bibinfo {author} {\bibfnamefont
  {L.}~\bibnamefont {Le~Guyader}}, \bibinfo {author} {\bibfnamefont
  {E.}~\bibnamefont {Mengotti}}, \bibinfo {author} {\bibfnamefont {L.~J.}\
  \bibnamefont {Heyderman}}, \bibinfo {author} {\bibfnamefont {F.}~\bibnamefont
  {Nolting}}, \bibinfo {author} {\bibfnamefont {A.}~\bibnamefont {Tsukamoto}},
  \bibinfo {author} {\bibfnamefont {A.}~\bibnamefont {Itoh}}, \bibinfo {author}
  {\bibfnamefont {D.}~\bibnamefont {Afanasiev}}, \bibinfo {author}
  {\bibfnamefont {B.~A.}\ \bibnamefont {Ivanov}}, \bibinfo {author}
  {\bibfnamefont {A.~M.}\ \bibnamefont {Kalashnikova}}, \bibinfo {author}
  {\bibfnamefont {K.}~\bibnamefont {Vahaplar}}, \bibinfo {author}
  {\bibfnamefont {J.}~\bibnamefont {Mentink}}, \bibinfo {author} {\bibfnamefont
  {A.}~\bibnamefont {Kirilyuk}}, \bibinfo {author} {\bibfnamefont
  {T.}~\bibnamefont {Rasing}}, \ and\ \bibinfo {author} {\bibfnamefont {A.~V.}\
  \bibnamefont {Kimel}},\ }\href {http://dx.doi.org/10.1038/ncomms1666}
  {\bibfield  {journal} {\bibinfo  {journal} {Nat Commun},\ }\textbf {\bibinfo
  {volume} {3}} (\bibinfo {year} {2012})}\BibitemShut {NoStop}%
\bibitem [{\citenamefont {Illg}\ \emph {et~al.}(2013)\citenamefont {Illg},
  \citenamefont {Haag},\ and\ \citenamefont {F{\"a}hnle}}]{illgPRB13}%
  \BibitemOpen
  \bibfield  {author} {\bibinfo {author} {\bibfnamefont {C.}~\bibnamefont
  {Illg}}, \bibinfo {author} {\bibfnamefont {M.}~\bibnamefont {Haag}}, \ and\
  \bibinfo {author} {\bibfnamefont {M.}~\bibnamefont {F{\"a}hnle}},\ }\Doi
  {10.1103/PhysRevB.88.214404} {\bibfield  {journal} {\bibinfo  {journal}
  {Phys. Rev. B},\ }\textbf {\bibinfo {volume} {88}},\ \bibinfo {pages}
  {214404} (\bibinfo {year} {2013})}\BibitemShut {NoStop}%
\bibitem [{\citenamefont {Zhang}\ \emph {et~al.}(2012)\citenamefont {Zhang},
  \citenamefont {Chuang}, \citenamefont {Zakeri},\ and\ \citenamefont
  {Kirschner}}]{zhangPRL12rt}%
  \BibitemOpen
  \bibfield  {author} {\bibinfo {author} {\bibfnamefont {Y.}~\bibnamefont
  {Zhang}}, \bibinfo {author} {\bibfnamefont {T.-H.}\ \bibnamefont {Chuang}},
  \bibinfo {author} {\bibfnamefont {K.}~\bibnamefont {Zakeri}}, \ and\ \bibinfo
  {author} {\bibfnamefont {J.}~\bibnamefont {Kirschner}},\ }\Doi
  {10.1103/PhysRevLett.109.087203} {\bibfield  {journal} {\bibinfo  {journal}
  {Phys. Rev. Lett.},\ }\textbf {\bibinfo {volume} {109}},\ \bibinfo {pages}
  {087203} (\bibinfo {year} {2012})}\BibitemShut {NoStop}%
\bibitem [{\citenamefont {Brataas}\ \emph {et~al.}(2000)\citenamefont
  {Brataas}, \citenamefont {Nazarov},\ and\ \citenamefont
  {Bauer}}]{brataasPRL00}%
  \BibitemOpen
  \bibfield  {author} {\bibinfo {author} {\bibfnamefont {A.}~\bibnamefont
  {Brataas}}, \bibinfo {author} {\bibfnamefont {Y.~V.}\ \bibnamefont
  {Nazarov}}, \ and\ \bibinfo {author} {\bibfnamefont {G.~E.~W.}\ \bibnamefont
  {Bauer}},\ }\href@noop {} {\bibfield  {journal} {\bibinfo  {journal} {Phys.
  Rev. Lett.},\ }\textbf {\bibinfo {volume} {84}},\ \bibinfo {pages} {2481}
  (\bibinfo {year} {2000})}\BibitemShut {NoStop}%
\bibitem [{\citenamefont {Tserkovnyak}\ \emph {et~al.}(2002)\citenamefont
  {Tserkovnyak}, \citenamefont {Brataas},\ and\ \citenamefont
  {Bauer}}]{tserkovPRL02sp}%
  \BibitemOpen
  \bibfield  {author} {\bibinfo {author} {\bibfnamefont {Y.}~\bibnamefont
  {Tserkovnyak}}, \bibinfo {author} {\bibfnamefont {A.}~\bibnamefont
  {Brataas}}, \ and\ \bibinfo {author} {\bibfnamefont {G.~E.~W.}\ \bibnamefont
  {Bauer}},\ }\href@noop {} {\bibfield  {journal} {\bibinfo  {journal} {Phys.
  Rev. Lett.},\ }\textbf {\bibinfo {volume} {88}},\ \bibinfo {eid} {117601}
  (\bibinfo {year} {2002})}\BibitemShut {NoStop}%
\bibitem [{\citenamefont {Tserkovnyak}\ \emph {et~al.}(2005)\citenamefont
  {Tserkovnyak}, \citenamefont {Brataas}, \citenamefont {Bauer},\ and\
  \citenamefont {Halperin}}]{tserkovRMP05}%
  \BibitemOpen
  \bibfield  {author} {\bibinfo {author} {\bibfnamefont {Y.}~\bibnamefont
  {Tserkovnyak}}, \bibinfo {author} {\bibfnamefont {A.}~\bibnamefont
  {Brataas}}, \bibinfo {author} {\bibfnamefont {G.~E.~W.}\ \bibnamefont
  {Bauer}}, \ and\ \bibinfo {author} {\bibfnamefont {B.~I.}\ \bibnamefont
  {Halperin}},\ }\href@noop {} {\bibfield  {journal} {\bibinfo  {journal} {Rev.
  Mod. Phys.},\ }\textbf {\bibinfo {volume} {77}},\ \bibinfo {eid} {1375}
  (\bibinfo {year} {2005})}\BibitemShut {NoStop}%
\bibitem [{\citenamefont {Hoffman}\ \emph {et~al.}(2013)\citenamefont
  {Hoffman}, \citenamefont {Sato},\ and\ \citenamefont
  {Tserkovnyak}}]{hoffmanPRB13}%
  \BibitemOpen
  \bibfield  {author} {\bibinfo {author} {\bibfnamefont {S.}~\bibnamefont
  {Hoffman}}, \bibinfo {author} {\bibfnamefont {K.}~\bibnamefont {Sato}}, \
  and\ \bibinfo {author} {\bibfnamefont {Y.}~\bibnamefont {Tserkovnyak}},\
  }\Doi {10.1103/PhysRevB.88.064408} {\bibfield  {journal} {\bibinfo  {journal}
  {Phys. Rev. B},\ }\textbf {\bibinfo {volume} {88}},\ \bibinfo {pages}
  {064408} (\bibinfo {year} {2013})}\BibitemShut {NoStop}%
\bibitem [{\citenamefont {Tserkovnyak}\ \emph {et~al.}(2009)\citenamefont
  {Tserkovnyak}, \citenamefont {Hankiewicz},\ and\ \citenamefont
  {Vignale}}]{tserkovPRB09sd}%
  \BibitemOpen
  \bibfield  {author} {\bibinfo {author} {\bibfnamefont {Y.}~\bibnamefont
  {Tserkovnyak}}, \bibinfo {author} {\bibfnamefont {E.~M.}\ \bibnamefont
  {Hankiewicz}}, \ and\ \bibinfo {author} {\bibfnamefont {G.}~\bibnamefont
  {Vignale}},\ }\Doi {10.1103/PhysRevB.79.094415} {\bibfield  {journal}
  {\bibinfo  {journal} {Phys. Rev. B},\ }\textbf {\bibinfo {volume} {79}},\
  \bibinfo {eid} {094415} (\bibinfo {year} {2009})}\BibitemShut {NoStop}%
\bibitem [{\citenamefont {Bender}\ \emph {et~al.}(2012)\citenamefont {Bender},
  \citenamefont {Duine},\ and\ \citenamefont {Tserkovnyak}}]{benderPRL12}%
  \BibitemOpen
  \bibfield  {author} {\bibinfo {author} {\bibfnamefont {S.~A.}\ \bibnamefont
  {Bender}}, \bibinfo {author} {\bibfnamefont {R.~A.}\ \bibnamefont {Duine}}, \
  and\ \bibinfo {author} {\bibfnamefont {Y.}~\bibnamefont {Tserkovnyak}},\
  }\Doi {10.1103/PhysRevLett.108.246601} {\bibfield  {journal} {\bibinfo
  {journal} {Phys. Rev. Lett.},\ }\textbf {\bibinfo {volume} {108}},\ \bibinfo
  {pages} {246601} (\bibinfo {year} {2012})}\BibitemShut {NoStop}%
\bibitem [{\citenamefont {Bender}\ \emph {et~al.}(2014)\citenamefont {Bender},
  \citenamefont {Duine}, \citenamefont {Brataas},\ and\ \citenamefont
  {Tserkovnyak}}]{benderPRB14}%
  \BibitemOpen
  \bibfield  {author} {\bibinfo {author} {\bibfnamefont {S.~A.}\ \bibnamefont
  {Bender}}, \bibinfo {author} {\bibfnamefont {R.~A.}\ \bibnamefont {Duine}},
  \bibinfo {author} {\bibfnamefont {A.}~\bibnamefont {Brataas}}, \ and\
  \bibinfo {author} {\bibfnamefont {Y.}~\bibnamefont {Tserkovnyak}},\ }\Doi
  {10.1103/PhysRevB.90.094409} {\bibfield  {journal} {\bibinfo  {journal}
  {Phys. Rev. B},\ }\textbf {\bibinfo {volume} {90}},\ \bibinfo {pages}
  {094409} (\bibinfo {year} {2014})}\BibitemShut {NoStop}%
\bibitem [{\citenamefont {La-O-Vorakiat}\ \emph {et~al.}(2012)\citenamefont
  {La-O-Vorakiat}, \citenamefont {Turgut}, \citenamefont {Teale}, \citenamefont
  {Kapteyn}, \citenamefont {Murnane}, \citenamefont {Mathias}, \citenamefont
  {Aeschlimann}, \citenamefont {Schneider}, \citenamefont {Shaw}, \citenamefont
  {Nembach},\ and\ \citenamefont {Silva}}]{vorakiatPRX12}%
  \BibitemOpen
  \bibfield  {author} {\bibinfo {author} {\bibfnamefont {C.}~\bibnamefont
  {La-O-Vorakiat}}, \bibinfo {author} {\bibfnamefont {E.}~\bibnamefont
  {Turgut}}, \bibinfo {author} {\bibfnamefont {C.~A.}\ \bibnamefont {Teale}},
  \bibinfo {author} {\bibfnamefont {H.~C.}\ \bibnamefont {Kapteyn}}, \bibinfo
  {author} {\bibfnamefont {M.~M.}\ \bibnamefont {Murnane}}, \bibinfo {author}
  {\bibfnamefont {S.}~\bibnamefont {Mathias}}, \bibinfo {author} {\bibfnamefont
  {M.}~\bibnamefont {Aeschlimann}}, \bibinfo {author} {\bibfnamefont {C.~M.}\
  \bibnamefont {Schneider}}, \bibinfo {author} {\bibfnamefont {J.~M.}\
  \bibnamefont {Shaw}}, \bibinfo {author} {\bibfnamefont {H.~T.}\ \bibnamefont
  {Nembach}}, \ and\ \bibinfo {author} {\bibfnamefont {T.~J.}\ \bibnamefont
  {Silva}},\ }\Doi {10.1103/PhysRevX.2.011005} {\bibfield  {journal} {\bibinfo
  {journal} {Phys. Rev. X},\ }\textbf {\bibinfo {volume} {2}},\ \bibinfo
  {pages} {011005} (\bibinfo {year} {2012})}\BibitemShut {NoStop}%
\bibitem [{\citenamefont {Mitchell}(1957)}]{mitchellPR57}%
  \BibitemOpen
  \bibfield  {author} {\bibinfo {author} {\bibfnamefont {A.~H.}\ \bibnamefont
  {Mitchell}},\ }\href@noop {} {\bibfield  {journal} {\bibinfo  {journal}
  {Phys. Rev.},\ }\textbf {\bibinfo {volume} {105}},\ \bibinfo {pages} {1439}
  (\bibinfo {year} {1957})}\BibitemShut {NoStop}%
\bibitem [{\citenamefont {Heinrich}\ \emph {et~al.}(1967)\citenamefont
  {Heinrich}, \citenamefont {Fraitov{\'a}},\ and\ \citenamefont
  {Kambersk{\'y}}}]{heinrichPSS67}%
  \BibitemOpen
  \bibfield  {author} {\bibinfo {author} {\bibfnamefont {B.}~\bibnamefont
  {Heinrich}}, \bibinfo {author} {\bibfnamefont {D.}~\bibnamefont
  {Fraitov{\'a}}}, \ and\ \bibinfo {author} {\bibfnamefont {V.}~\bibnamefont
  {Kambersk{\'y}}},\ }\href@noop {} {\bibfield  {journal} {\bibinfo  {journal}
  {Phys. Status Solidi},\ }\textbf {\bibinfo {volume} {23}},\ \bibinfo {pages}
  {501} (\bibinfo {year} {1967})}\BibitemShut {NoStop}%
\bibitem [{\citenamefont {Tserkovnyak}\ \emph {et~al.}(2004)\citenamefont
  {Tserkovnyak}, \citenamefont {Fiete},\ and\ \citenamefont
  {Halperin}}]{tserkovAPL04}%
  \BibitemOpen
  \bibfield  {author} {\bibinfo {author} {\bibfnamefont {Y.}~\bibnamefont
  {Tserkovnyak}}, \bibinfo {author} {\bibfnamefont {G.~A.}\ \bibnamefont
  {Fiete}}, \ and\ \bibinfo {author} {\bibfnamefont {B.~I.}\ \bibnamefont
  {Halperin}},\ }\href@noop {} {\bibfield  {journal} {\bibinfo  {journal}
  {Appl. Phys. Lett.},\ }\textbf {\bibinfo {volume} {84}},\ \bibinfo {pages}
  {5234} (\bibinfo {year} {2004})}\BibitemShut {NoStop}%
\bibitem [{\citenamefont {Mueller}\ \emph {et~al.}(2013)\citenamefont
  {Mueller}, \citenamefont {Baral}, \citenamefont {Vollmar}, \citenamefont
  {Cinchetti}, \citenamefont {Aeschlimann}, \citenamefont {Schneider},\ and\
  \citenamefont {Rethfeld}}]{PhysRevLett.111.167204}%
  \BibitemOpen
  \bibfield  {author} {\bibinfo {author} {\bibfnamefont {B.~Y.}\ \bibnamefont
  {Mueller}}, \bibinfo {author} {\bibfnamefont {A.}~\bibnamefont {Baral}},
  \bibinfo {author} {\bibfnamefont {S.}~\bibnamefont {Vollmar}}, \bibinfo
  {author} {\bibfnamefont {M.}~\bibnamefont {Cinchetti}}, \bibinfo {author}
  {\bibfnamefont {M.}~\bibnamefont {Aeschlimann}}, \bibinfo {author}
  {\bibfnamefont {H.~C.}\ \bibnamefont {Schneider}}, \ and\ \bibinfo {author}
  {\bibfnamefont {B.}~\bibnamefont {Rethfeld}},\ }\Doi
  {10.1103/PhysRevLett.111.167204} {\bibfield  {journal} {\bibinfo  {journal}
  {Phys. Rev. Lett.},\ }\textbf {\bibinfo {volume} {111}},\ \bibinfo {pages}
  {167204} (\bibinfo {year} {2013})}\BibitemShut {NoStop}%
\bibitem [{\citenamefont {Barker}\ \emph {et~al.}(2013)\citenamefont {Barker},
  \citenamefont {Atxitia}, \citenamefont {Ostler}, \citenamefont {Hovorka},
  \citenamefont {Chubykalo-Fesenko},\ and\ \citenamefont
  {Chantrell}}]{Barker:2013kx}%
  \BibitemOpen
  \bibfield  {author} {\bibinfo {author} {\bibfnamefont {J.}~\bibnamefont
  {Barker}}, \bibinfo {author} {\bibfnamefont {U.}~\bibnamefont {Atxitia}},
  \bibinfo {author} {\bibfnamefont {T.~A.}\ \bibnamefont {Ostler}}, \bibinfo
  {author} {\bibfnamefont {O.}~\bibnamefont {Hovorka}}, \bibinfo {author}
  {\bibfnamefont {O.}~\bibnamefont {Chubykalo-Fesenko}}, \ and\ \bibinfo
  {author} {\bibfnamefont {R.~W.}\ \bibnamefont {Chantrell}},\ }\href
  {http://dx.doi.org/10.1038/srep03262} {\bibfield  {journal} {\bibinfo
  {journal} {Sci. Rep.},\ }\textbf {\bibinfo {volume} {3}} (\bibinfo {year}
  {2013})}\BibitemShut {NoStop}%
\bibitem [{\citenamefont {Manchon}\ \emph {et~al.}(2012)\citenamefont
  {Manchon}, \citenamefont {Li}, \citenamefont {Xu},\ and\ \citenamefont
  {Zhang}}]{manchonPRB12}%
  \BibitemOpen
  \bibfield  {author} {\bibinfo {author} {\bibfnamefont {A.}~\bibnamefont
  {Manchon}}, \bibinfo {author} {\bibfnamefont {Q.}~\bibnamefont {Li}},
  \bibinfo {author} {\bibfnamefont {L.}~\bibnamefont {Xu}}, \ and\ \bibinfo
  {author} {\bibfnamefont {S.}~\bibnamefont {Zhang}},\ }\Doi
  {10.1103/PhysRevB.85.064408} {\bibfield  {journal} {\bibinfo  {journal}
  {Phys. Rev. B},\ }\textbf {\bibinfo {volume} {85}},\ \bibinfo {pages}
  {064408} (\bibinfo {year} {2012})}\BibitemShut {NoStop}%
\bibitem [{\citenamefont {Meservey}\ and\ \citenamefont
  {Tedrow}(1978)}]{meserveyPRL78}%
  \BibitemOpen
  \bibfield  {author} {\bibinfo {author} {\bibfnamefont {R.}~\bibnamefont
  {Meservey}}\ and\ \bibinfo {author} {\bibfnamefont {P.~M.}\ \bibnamefont
  {Tedrow}},\ }\Doi {10.1103/PhysRevLett.41.805} {\bibfield  {journal}
  {\bibinfo  {journal} {Phys. Rev. Lett.},\ }\textbf {\bibinfo {volume} {41}},\
  \bibinfo {pages} {805} (\bibinfo {year} {1978})}\BibitemShut {NoStop}%
\bibitem [{\citenamefont {Garanin}(1997)}]{garaninPRB97}%
  \BibitemOpen
  \bibfield  {author} {\bibinfo {author} {\bibfnamefont {D.~A.}\ \bibnamefont
  {Garanin}},\ }\Doi {10.1103/PhysRevB.55.3050} {\bibfield  {journal} {\bibinfo
   {journal} {Phys. Rev. B},\ }\textbf {\bibinfo {volume} {55}},\ \bibinfo
  {pages} {3050} (\bibinfo {year} {1997})}\BibitemShut {NoStop}%
\bibitem [{\citenamefont {Atxitia}\ \emph {et~al.}(2010)\citenamefont
  {Atxitia}, \citenamefont {Chubykalo-Fesenko}, \citenamefont {Walowski},
  \citenamefont {Mann},\ and\ \citenamefont
  {M\"unzenberg}}]{PhysRevB.81.174401}%
  \BibitemOpen
  \bibfield  {author} {\bibinfo {author} {\bibfnamefont {U.}~\bibnamefont
  {Atxitia}}, \bibinfo {author} {\bibfnamefont {O.}~\bibnamefont
  {Chubykalo-Fesenko}}, \bibinfo {author} {\bibfnamefont {J.}~\bibnamefont
  {Walowski}}, \bibinfo {author} {\bibfnamefont {A.}~\bibnamefont {Mann}}, \
  and\ \bibinfo {author} {\bibfnamefont {M.}~\bibnamefont {M\"unzenberg}},\
  }\Doi {10.1103/PhysRevB.81.174401} {\bibfield  {journal} {\bibinfo  {journal}
  {Phys. Rev. B},\ }\textbf {\bibinfo {volume} {81}},\ \bibinfo {pages}
  {174401} (\bibinfo {year} {2010})}\BibitemShut {NoStop}%
\bibitem [{\citenamefont {Brown}(1963)}]{brownPR63}%
  \BibitemOpen
  \bibfield  {author} {\bibinfo {author} {\bibfnamefont {W.~F.}\ \bibnamefont
  {Brown}},\ }\href@noop {} {\bibfield  {journal} {\bibinfo  {journal} {Phys.
  Rev.},\ }\textbf {\bibinfo {volume} {130}},\ \bibinfo {pages} {1677}
  (\bibinfo {year} {1963})}\BibitemShut {NoStop}%
\bibitem [{\citenamefont {Kubo}\ and\ \citenamefont
  {Hashitsume}(1970)}]{Kubo01061970}%
  \BibitemOpen
  \bibfield  {author} {\bibinfo {author} {\bibfnamefont {R.}~\bibnamefont
  {Kubo}}\ and\ \bibinfo {author} {\bibfnamefont {N.}~\bibnamefont
  {Hashitsume}},\ }\Doi {10.1143/PTPS.46.210} {\bibfield  {journal} {\bibinfo
  {journal} {Progress of Theoretical Physics Supplement},\ }\textbf {\bibinfo
  {volume} {46}},\ \bibinfo {pages} {210} (\bibinfo {year} {1970})}\BibitemShut
  {NoStop}%
\bibitem [{\citenamefont {Kampfrath}\ \emph {et~al.}(2002)\citenamefont
  {Kampfrath}, \citenamefont {Ulbrich}, \citenamefont {Leuenberger},
  \citenamefont {M\"unzenberg}, \citenamefont {Sass},\ and\ \citenamefont
  {Felsch}}]{PhysRevB.65.104429}%
  \BibitemOpen
  \bibfield  {author} {\bibinfo {author} {\bibfnamefont {T.}~\bibnamefont
  {Kampfrath}}, \bibinfo {author} {\bibfnamefont {R.~G.}\ \bibnamefont
  {Ulbrich}}, \bibinfo {author} {\bibfnamefont {F.}~\bibnamefont
  {Leuenberger}}, \bibinfo {author} {\bibfnamefont {M.}~\bibnamefont
  {M\"unzenberg}}, \bibinfo {author} {\bibfnamefont {B.}~\bibnamefont {Sass}},
  \ and\ \bibinfo {author} {\bibfnamefont {W.}~\bibnamefont {Felsch}},\ }\Doi
  {10.1103/PhysRevB.65.104429} {\bibfield  {journal} {\bibinfo  {journal}
  {Phys. Rev. B},\ }\textbf {\bibinfo {volume} {65}},\ \bibinfo {pages}
  {104429} (\bibinfo {year} {2002})}\BibitemShut {NoStop}%
\bibitem [{\citenamefont {Carpene}\ \emph {et~al.}(2008)\citenamefont
  {Carpene}, \citenamefont {Mancini}, \citenamefont {Dallera}, \citenamefont
  {Brenna}, \citenamefont {Puppin},\ and\ \citenamefont
  {De~Silvestri}}]{PhysRevB.78.174422}%
  \BibitemOpen
  \bibfield  {author} {\bibinfo {author} {\bibfnamefont {E.}~\bibnamefont
  {Carpene}}, \bibinfo {author} {\bibfnamefont {E.}~\bibnamefont {Mancini}},
  \bibinfo {author} {\bibfnamefont {C.}~\bibnamefont {Dallera}}, \bibinfo
  {author} {\bibfnamefont {M.}~\bibnamefont {Brenna}}, \bibinfo {author}
  {\bibfnamefont {E.}~\bibnamefont {Puppin}}, \ and\ \bibinfo {author}
  {\bibfnamefont {S.}~\bibnamefont {De~Silvestri}},\ }\Doi
  {10.1103/PhysRevB.78.174422} {\bibfield  {journal} {\bibinfo  {journal}
  {Phys. Rev. B},\ }\textbf {\bibinfo {volume} {78}},\ \bibinfo {pages}
  {174422} (\bibinfo {year} {2008})}\BibitemShut {NoStop}%
\bibitem [{\citenamefont {Weber}\ \emph {et~al.}(2011)\citenamefont {Weber},
  \citenamefont {Pressacco}, \citenamefont {G\"unther}, \citenamefont
  {Mancini}, \citenamefont {Oppeneer},\ and\ \citenamefont
  {Back}}]{PhysRevB.84.132412}%
  \BibitemOpen
  \bibfield  {author} {\bibinfo {author} {\bibfnamefont {A.}~\bibnamefont
  {Weber}}, \bibinfo {author} {\bibfnamefont {F.}~\bibnamefont {Pressacco}},
  \bibinfo {author} {\bibfnamefont {S.}~\bibnamefont {G\"unther}}, \bibinfo
  {author} {\bibfnamefont {E.}~\bibnamefont {Mancini}}, \bibinfo {author}
  {\bibfnamefont {P.~M.}\ \bibnamefont {Oppeneer}}, \ and\ \bibinfo {author}
  {\bibfnamefont {C.~H.}\ \bibnamefont {Back}},\ }\Doi
  {10.1103/PhysRevB.84.132412} {\bibfield  {journal} {\bibinfo  {journal}
  {Phys. Rev. B},\ }\textbf {\bibinfo {volume} {84}},\ \bibinfo {pages}
  {132412} (\bibinfo {year} {2011})}\BibitemShut {NoStop}%
\bibitem [{\citenamefont {Mathias}\ \emph {et~al.}(2012)\citenamefont
  {Mathias}, \citenamefont {La-O-Vorakiat}, \citenamefont {Grychtol},
  \citenamefont {Granitzka}, \citenamefont {Turgut}, \citenamefont {Shaw},
  \citenamefont {Adam}, \citenamefont {Nembach}, \citenamefont {Siemens},
  \citenamefont {Eich}, \citenamefont {Schneider}, \citenamefont {Silva},
  \citenamefont {Aeschlimann}, \citenamefont {Murnane},\ and\ \citenamefont
  {Kapteyn}}]{mathiasPNAS12}%
  \BibitemOpen
  \bibfield  {author} {\bibinfo {author} {\bibfnamefont {S.}~\bibnamefont
  {Mathias}}, \bibinfo {author} {\bibfnamefont {C.}~\bibnamefont
  {La-O-Vorakiat}}, \bibinfo {author} {\bibfnamefont {P.}~\bibnamefont
  {Grychtol}}, \bibinfo {author} {\bibfnamefont {P.}~\bibnamefont {Granitzka}},
  \bibinfo {author} {\bibfnamefont {E.}~\bibnamefont {Turgut}}, \bibinfo
  {author} {\bibfnamefont {J.~M.}\ \bibnamefont {Shaw}}, \bibinfo {author}
  {\bibfnamefont {R.}~\bibnamefont {Adam}}, \bibinfo {author} {\bibfnamefont
  {H.~T.}\ \bibnamefont {Nembach}}, \bibinfo {author} {\bibfnamefont {M.~E.}\
  \bibnamefont {Siemens}}, \bibinfo {author} {\bibfnamefont {S.}~\bibnamefont
  {Eich}}, \bibinfo {author} {\bibfnamefont {C.~M.}\ \bibnamefont {Schneider}},
  \bibinfo {author} {\bibfnamefont {T.~J.}\ \bibnamefont {Silva}}, \bibinfo
  {author} {\bibfnamefont {M.}~\bibnamefont {Aeschlimann}}, \bibinfo {author}
  {\bibfnamefont {M.~M.}\ \bibnamefont {Murnane}}, \ and\ \bibinfo {author}
  {\bibfnamefont {H.~C.}\ \bibnamefont {Kapteyn}},\ }\href@noop {} {\bibfield
  {journal} {\bibinfo  {journal} {Proc. Natl. Acad. Sci. USA},\ }\textbf
  {\bibinfo {volume} {109}},\ \bibinfo {pages} {4792} (\bibinfo {year}
  {2012})}\BibitemShut {NoStop}%
\bibitem [{\citenamefont {Hatami}\ \emph {et~al.}(2007)\citenamefont {Hatami},
  \citenamefont {Bauer}, \citenamefont {Zhang},\ and\ \citenamefont
  {Kelly}}]{PhysRevLett.99.066603}%
  \BibitemOpen
  \bibfield  {author} {\bibinfo {author} {\bibfnamefont {M.}~\bibnamefont
  {Hatami}}, \bibinfo {author} {\bibfnamefont {G.~E.~W.}\ \bibnamefont
  {Bauer}}, \bibinfo {author} {\bibfnamefont {Q.}~\bibnamefont {Zhang}}, \ and\
  \bibinfo {author} {\bibfnamefont {P.~J.}\ \bibnamefont {Kelly}},\ }\Doi
  {10.1103/PhysRevLett.99.066603} {\bibfield  {journal} {\bibinfo  {journal}
  {Phys. Rev. Lett.},\ }\textbf {\bibinfo {volume} {99}},\ \bibinfo {pages}
  {066603} (\bibinfo {year} {2007})}\BibitemShut {NoStop}%
\bibitem [{\citenamefont {Mizukami}\ \emph {et~al.}(2001)\citenamefont
  {Mizukami}, \citenamefont {Ando},\ and\ \citenamefont
  {Miyazaki}}]{mizukamiJJAP01}%
  \BibitemOpen
  \bibfield  {author} {\bibinfo {author} {\bibfnamefont {S.}~\bibnamefont
  {Mizukami}}, \bibinfo {author} {\bibfnamefont {Y.}~\bibnamefont {Ando}}, \
  and\ \bibinfo {author} {\bibfnamefont {T.}~\bibnamefont {Miyazaki}},\
  }\href@noop {} {\bibfield  {journal} {\bibinfo  {journal} {Jpn. J. Appl.
  Phys.},\ }\textbf {\bibinfo {volume} {40}},\ \bibinfo {pages} {580} (\bibinfo
  {year} {2001})}\BibitemShut {NoStop}%
\bibitem [{\citenamefont {Mizukami}\ \emph {et~al.}(2002)\citenamefont
  {Mizukami}, \citenamefont {Ando},\ and\ \citenamefont
  {Miyazaki}}]{mizukamiPRB02}%
  \BibitemOpen
  \bibfield  {author} {\bibinfo {author} {\bibfnamefont {S.}~\bibnamefont
  {Mizukami}}, \bibinfo {author} {\bibfnamefont {Y.}~\bibnamefont {Ando}}, \
  and\ \bibinfo {author} {\bibfnamefont {T.}~\bibnamefont {Miyazaki}},\
  }\href@noop {} {\bibfield  {journal} {\bibinfo  {journal} {Phys. Rev. B},\
  }\textbf {\bibinfo {volume} {66}},\ \bibinfo {eid} {104413} (\bibinfo {year}
  {2002})}\BibitemShut {NoStop}%
\bibitem [{\citenamefont {Liu}\ \emph {et~al.}(2014)\citenamefont {Liu},
  \citenamefont {Yuan}, \citenamefont {Wesselink}, \citenamefont {Starikov},\
  and\ \citenamefont {Kelly}}]{PhysRevLett.113.207202}%
  \BibitemOpen
  \bibfield  {author} {\bibinfo {author} {\bibfnamefont {Y.}~\bibnamefont
  {Liu}}, \bibinfo {author} {\bibfnamefont {Z.}~\bibnamefont {Yuan}}, \bibinfo
  {author} {\bibfnamefont {J.~H.}\ \bibnamefont {Wesselink}, \bibfnamefont
  {R.}}, \bibinfo {author} {\bibfnamefont {A.~A.}\ \bibnamefont {Starikov}}, \
  and\ \bibinfo {author} {\bibfnamefont {P.~J.}\ \bibnamefont {Kelly}},\ }\Doi
  {10.1103/PhysRevLett.113.207202} {\bibfield  {journal} {\bibinfo  {journal}
  {Phys. Rev. Lett.},\ }\textbf {\bibinfo {volume} {113}},\ \bibinfo {pages}
  {207202} (\bibinfo {year} {2014})}\BibitemShut {NoStop}%
\end{thebibliography}
\end{document}